\documentclass[longauth]{aa}
\usepackage{graphicx}
\usepackage{txfonts}
\usepackage{natbib}
\usepackage{tabularx}
\bibpunct{(}{)}{;}{a}{}{,}

\def\beq{\begin{equation}}
\def\eeq{\end{equation}}

\begin{document}

\title{The VIMOS Public Extragalactic Survey (VIPERS) \thanks{based on
observations collected at the European Southern Observatory, Cerro Paranal,
Chile, using the Very Large Telescope under programs 182.A-0886 and partly
070.A-9007. Also based on observations obtained with MegaPrime/MegaCam, a joint
project of CFHT and CEA/DAPNIA, at the Canada-France-Hawaii Telescope (CFHT),
which is operated by the National Research Council (NRC) of Canada, the Institut
National des Sciences de l’Univers of the Centre National de la Recherche
Scientifique (CNRS) of France, and the University of Hawaii. This work is based
in part on data products produced at TERAPIX and the Canadian Astronomy Data
Centre as part of the Canada-France-Hawaii Telescope Legacy Survey, a
collaborative project of NRC and CNRS. The VIPERS web site is
{\tt http://vipers.inaf.it/}.}}
\subtitle{\bf First Data Release of 57\,204 spectroscopic measurements}

\author{
B. Garilli \inst{1}
\and L.~Guzzo\inst{2,3}
\and M.~Scodeggio\inst{1} 
%\and the VIPERS Team
\and M.~Bolzonella\inst{9}           
% group 2
\and U.~Abbas\inst{5}
\and C.~Adami\inst{4}
\and S.~Arnouts\inst{6,4}
\and J.~Bel\inst{7}
\and D.~Bottini\inst{1}
\and E.~Branchini\inst{10,27,28}
\and A.~Cappi\inst{9}
\and J.~Coupon\inst{12}
\and O.~Cucciati\inst{9,17}           
\and I.~Davidzon\inst{9,17}
\and G.~De Lucia\inst{13}
\and S.~de la Torre\inst{14}
\and P.~Franzetti\inst{1}
\and A.~Fritz\inst{1}
\and M.~Fumana\inst{1}
\and B.~R.~Granett\inst{2}
\and O.~Ilbert\inst{4}
\and A.~Iovino\inst{2}
\and J.~Krywult\inst{15}
\and V.~Le Brun\inst{4}
\and O.~Le F\`evre\inst{4}
\and D.~Maccagni\inst{1}
\and K.~Ma{\l}ek\inst{16}
\and F.~Marulli\inst{17,18,9}
\and H.~J.~McCracken\inst{19}
\and L.~Paioro\inst{1}
\and M.~Polletta\inst{1}
\and A.~Pollo\inst{21,22}
\and H.~Schlagenhaufer\inst{23,20}
\and L.~A.~M.~Tasca\inst{4}
\and R.~Tojeiro\inst{11}
\and D.~Vergani\inst{24}
\and G.~Zamorani\inst{9}
\and A.~Zanichelli\inst{25}
\and A.~Burden\inst{11}
\and C.~Di Porto\inst{9}
\and A.~Marchetti\inst{2,26} 
\and C.~Marinoni\inst{7}
\and Y.~Mellier\inst{19}
\and L.~Moscardini\inst{17,18,9}
\and R.~C.~Nichol\inst{11}
\and J.~A.~Peacock\inst{14}
\and W.~J.~Percival\inst{11}
\and S.~Phleps\inst{20}
\and M.~Wolk\inst{19}
}
   
\institute{
INAF - Istituto di Astrofisica Spaziale e Fisica Cosmica Milano, via Bassini 15,
20133 Milano, Italy%1
\and INAF - Osservatorio Astronomico di Brera, Via Brera 28, 20122 Milano, via
E. Bianchi 46, 23807 Merate, Italy %2
\and Dipartimento di Fisica, Universit\`a di Milano-Bicocca, P.zza della Scienza
3, I-20126 Milano, Italy %3
\and Aix Marseille Universit\'e, CNRS, LAM (Laboratoire d'Astrophysique de
Marseille) UMR 7326, 13388, Marseille, France  %4
\and INAF - Osservatorio Astrofisico di Torino, 10025 Pino Torinese, Italy %5
\and Canada-France-Hawaii Telescope, 65--1238 Mamalahoa Highway, Kamuela, HI
96743, USA %6
\and Aix-Marseille Universit\'e, CNRS, CPT (Centre de Physique  Th\'eorique) UMR
7332, F-13288 Marseille, France %7
\and Universit\'{e} de Lyon, F-69003 Lyon, France %8
\and INAF - Osservatorio Astronomico di Bologna, via Ranzani 1, I-40127,
Bologna, Italy %9
\and Dipartimento di Matematica e Fisica, Universit\`{a} degli Studi Roma Tre,
via della Vasca Navale 84, 00146 Roma, Italy %10
\and Institute of Cosmology and Gravitation, Dennis Sciama Building, University
of Portsmouth, Burnaby Road, Portsmouth, PO1 3FX %11
\and Institute of Astronomy and Astrophysics, Academia Sinica, P.O. Box 23-141,
Taipei 10617, Taiwan%12
\and INAF - Osservatorio Astronomico di Trieste, via G. B. Tiepolo 11, 34143
Trieste, Italy %13
\and SUPA, Institute for Astronomy, University of Edinburgh, Royal Observatory,
Blackford Hill, Edinburgh EH9 3HJ, UK %14
\and Institute of Physics, Jan Kochanowski University, ul. Swietokrzyska 15,
25-406 Kielce, Poland %15
\and Department of Particle and Astrophysical Science, Nagoya University,
Furo-cho, Chikusa-ku, 464-8602 Nagoya, Japan %16
\and Dipartimento di Fisica e Astronomia - Universit\`{a} di Bologna, viale
Berti Pichat 6/2, I-40127 Bologna, Italy %17
\and INFN, Sezione di Bologna, viale Berti Pichat 6/2, I-40127 Bologna, Italy
%18
\and Institute d'Astrophysique de Paris, UMR7095 CNRS, Universit\'{e} Pierre et
Marie Curie, 98 bis Boulevard Arago, 75014 Paris, France %19
\and Max-Planck-Institut f\"{u}r Extraterrestrische Physik, D-84571 Garching b.
M\"{u}nchen, Germany %20
\and Astronomical Observatory of the Jagiellonian University, Orla 171, 30-001
Cracow, Poland %21
\and National Centre for Nuclear Research, ul. Hoza 69, 00-681 Warszawa, Poland
%22
\and Universit\"{a}tssternwarte M\"{u}nchen, Ludwig-Maximillians
Universit\"{a}t, Scheinerstr. 1, D-81679 M\"{u}nchen, Germany %23
\and INAF - Istituto di Astrofisica Spaziale e Fisica Cosmica Bologna, via
Gobetti 101, I-40129 Bologna, Italy %24
\and INAF - Istituto di Radioastronomia, via Gobetti 101, I-40129, Bologna,
Italy %25
\and Universit\`{a} degli Studi di Milano, via G. Celoria 16, 20130 Milano,
Italy %26
\and INFN, Sezione di Roma Tre, via della Vasca Navale 84, I-00146 Roma, Italy
%27
\and INAF - Osservatorio Astronomico di Roma, via Frascati 33, I-00040 Monte
Porzio Catone (RM), Italy %28
}

\offprints{B.Garilli, bianca@lambrate.inaf.it}

\date{Accepted for publication in A\&A}

% \abstract{}{}{}{}{} 
% 5 {} token are mandatory
 
\abstract 
{We present the first Public Data Release (PDR-1) of the VIMOS Public Extragalactic Survey (VIPERS). 
It comprises 57\,204 
spectroscopic measurements together with all additional
information necessary for optimal scientific exploitation of the data,
in particular the associated photometric measurements and quantification of the 
photometric and survey completeness.  VIPERS is an ESO Large Programme
 designed to build a spectroscopic sample of $\simeq 100\,000$  galaxies
 with $i_{\rm AB}<22.5$ and $0.5<z<1.2$ with high sampling rate ($\simeq
 45\%$).  The survey spectroscopic targets are selected from the
 CFHTLS-Wide five-band catalogues in the W1 and W4 fields.  The final
 survey will cover a total area of nearly $24$~deg$^2$, for a total
 comoving volume between $z=0.5$ and 1.2 of 
 $\simeq 4\times 10^7 \,h^{-3}{\rm Mpc}^3$ and a median galaxy redshift
 of $z\simeq 0.8$. 
The release presented in this paper 
includes data from virtually the entire
 W4 field and nearly half of the W1 
area, thus representing  64\% of the final dataset.  
We %present 
provide
a detailed description of sample selection, 
observations and data reduction procedures; 
we summarise  the global properties of the spectroscopic catalogue
and explain the associated data products and their use, 
and provide all the details for accessing the data through the survey
database ({\tt http://vipers.inaf.it}) where all information can be
queried interactively. 
} 

 
   


\authorrunning{B.Garilli et al.}
\titlerunning{VIPERS- First Public Data Release}
\maketitle

\section{Introduction}
The large-scale 
distribution of galaxies contains unique information on the structure of 
our Universe and the fundamental parameters of the cosmological model.
The relation of galaxy properties to large-scale structure in turn
provides important clues on the physics of galaxy formation within the
standard paradigm in which baryons are assembled inside dark-matter
halos \citep[e.g.][]{white_rees}.  Redshift surveys of the ``local''
($z<0.2$) Universe such as the 2dFGRS \citep{2df} and SDSS \citep{sdss}
include more than a million objects
observed over several thousand square degrees.
Thanks to such excellent statistics, large-scale structure studies
have been extended well into the linear regime ($r\gg 10 $ h$^{-1}$ Mpc) 
while 
at the same time having a detailed characterization of small-scale
clustering and its dependence on galaxy properties like luminosity, colour and
morphology  \citep[e.g.][]{2df_clus_type,2df_clus2,sdss_clus1,sdss_clus2}.
All these features and properties are expected to depend on redshift,
and different evolutionary paths can lead to similar observational
properties in the local universe. 
Ideally, one would 
like to be able to gather similarly large samples over comparably large 
volumes, at cosmologically relevant distances ($z>>0.3$). 

Pioneering deep surveys capable of measuring the evolution of
clustering since $z\sim 1$ date back to the
1990s and were
limited to a few hundred square arcminutes
\citep[e.g.][]{cfrs_clustering,cnoc_clustering}.
Studies extending further in redshift were limited to
specific color-selected samples based on the Lyman-break technique \citep[e.g.][]{steidel98}.
More recent surveys like GOODS \citep[e.g.][]{goods} 
%and DEEP \citep[e.g.][]{deep}
provided a broader view of the high-redshift population, but
still limited to small fields.
Only 
with the advent of multi-object spectrographs mounted on 10-m class telescopes,  
significant clustering studies of the general galaxy population at $z\sim
1$ became feasible as notably exploited by the
VVDS \citep{vvds_main} and the DEEP2 \citep{deep2} surveys,
followed by the zCOSMOS follow-up of the COSMOS HST field
\citep{zcosmos_10k}. 
While important clustering studies at $z\sim
1$ were produced, it became soon clear that these samples
remained in general dominated by field-to-field fluctuations (cosmic variance), 
as dramatically shown by the discrepancy between the VVDS and zCOSMOS
correlation functions  at $0.5<z<1$ \citep{delatorre10}.
Only the Wide extension of VVDS
   \citep{vvds_wide}, started to probe sufficient volume 
at these epochs, as to attempt cosmologically meaningful computations
\citep{guzzo_nature}, albeit still with large error bars.

Following those efforts, and somewhat complementarily, new generations
of cosmological surveys have mostly focused on 
   covering the largest possible volumes at intermediate depths, utilizing
   relatively low-density tracers, with the main goal of measuring the BAO signal
   at redshifts 0.4-0.8.
This is the case with the SDSS-3 BOSS project
   \citep{eisenstein11}, which extends the concept pioneered by the SDSS
   selection of Luminous Red Galaxies
   \citep[e.g.][]{anderson12, reid12}.  Similarly, the WiggleZ survey
   targets emission-line galaxies selected from UV observations of the 
   GALEX satellite \citep{wigglez, blake11, blake11b}. Both these
   surveys are characterized by a very large volume 
  ($1-2\, h^{-3}{\rm Gpc}^3$), and a relatively sparse galaxy population ($\sim
   10^{-4}\,h^{3}{\rm Mpc}^{-3}$).  
Complementarily, the GAMA survey
   \citep{gama} aims to achieve a number of redshifts similar to the 2dFGRS 
($\sim 200,000$) down to $r<19.8$, covering a
smaller volume ($z\simeq 0.5$) but with a sampling close to unity. 

The VIMOS Public Extragalactic Redshift Survey (VIPERS;
\citealt{vipers_main}), of which we present here the first public data
release, has been designed to collect $\sim 10^5$ redshifts
to the same depth of VVDS-Wide and zCOSMOS
 ($i_{\rm AB}<22.5$), but over a
significantly larger volume and with high sampling (${\rm
  \sim 40\%}$ ).   
The general aim of the project is to build a sample of the global galaxy population 
that matches in several respects those available locally ($z<0.2$) from the  2dFGRS  and SDSS  
projects, thus allowing combined evolutionary studies of both clustering and galaxy physical properties, 
on a comparable statistical footing. Building upon the experience and results of previous 
VIMOS surveys  VIPERS 
arguably provides the most detailed and representative picture
to date of the whole galaxy population and its large-scale structures, when the
Universe was about half its current age.  

The main goals of VIPERS are:

1.~To measure the clustering of galaxies at $\langle z\rangle \simeq 0.8$ on scales up
to $\sim 100~{\rm Mpc}~h^{-1}$, in order to: \\
(a)~extract cosmological information from
the large-scale shape of the power spectrum \citep{Bel2013, Granett2012,
  Xia2012};\\
(b)~quantify the dependence of galaxy clustering 
on galaxy physical properties
(such as luminosity and stellar mass) and its evolution with time \citep{vipers_clus_lum};\\
(c)~quantify the mean galaxy occupation of dark matter halos using small/intermediate-scale 
clustering and its time evolution;
% (de la Torre et al., in preparation); \\
(d)~study the halo occupation distribution (HOD) using
a combination of galaxy-galaxy lensing, galaxy clustering and stellar mass 
functions (Coupon et al., in preparation);\\ 
(e)~measure higher-order clustering statistics and characterize the
non-linear development of clustering over the past 7 billion years.

2.~To measure, at the same redshifts, the growth rate of structure using redshift-space
distortions in the observed clustering \citep{vipers_clus},
in particular exploiting the broad population and high spatial sampling of VIPERS
through the use of multiple tracers of the underlying mass density field.
% (Granett et al., in preparation).

3.~To precisely characterise the galaxy population at $\langle z\rangle \simeq 0.8$ in terms of
the distributions of fundamental properties such as luminosity,
colours and stellar mass, tracing their evolution with cosmic time
\citep{vipers_mf, vipers_cm}. 

4. To reconstruct the density field at $\langle z\rangle \simeq 0.8$ in order to:\\
(a)~determine the bias parameter and its evolution $b(z)$ (Branchini et al., in preparation);\\
(b) quantify at these redshifts the relationship between galaxy properties and local environment,
elucidating the role of mass and environment in the evolution of
galaxies. 
%(Cucciati et al., in preparation, Iovino et al., in preparation). 

5.~To provide the community with an unprecedented spectroscopic
database at $0.5<z<1.2$, including extensive information on galaxy
physical properties. The latter is made possible by combining the spectral information
 with the CFHTLS five-band magnitudes on which the survey
is based{\footnote{{\tt http://www.cfht.hawaii.edu/Science/CFHTLS/}}} 
\citep{CFHTLS}, plus additional ancillary data in the UV and infrared bands
(GALEX, UKIDSS, VISTA, SWIRE, VLA, XMM-LSS), 
enabling to derive Spectral Energy Distribution information (\citealt{vipers_pca}, \citealt{vipers_mf})
and automatic galaxy/AGN/stellar classification \citep{vipers_svm}.

Although carried out within
the standard (proprietary) scheme of the 
ESO Large Programmes, VIPERS was conceived from its start as a
public survey, with the clear idea that the range of science that
will be performed by the community should greatly exceed the core analyses
described above.  The set of data that
is described in this paper and made public with the VIPERS Public Data
Release 1 (PDR-1), is the same used for all papers of the first VIPERS
science release of March 2013. Several
aspects of the survey construction and the data are also discussed in
\citet{vipers_main} and are only briefly summarised here. 

The layout of the paper is as follows: \S2
summarises the survey strategy and design;
\S3 describes the VLT-VIMOS observations; \S4 discusses
the data reduction, including redshift estimation and quality tests;
\S5 presents the PDR-1 sample, discussing redshift errors and
comparison to external data; 
\S6 describes the survey
masks and discusses in detail the selection effects that
need to be taken into account for a proper use of the PDR-1 data; 
\S7
provides a first look at the VIPERS spectra, including a brief preliminary 
discussion of trends and correlations spectral features and galaxy classification schemes;
\S8 provides information on the data access in the VIPERS data base;
finally, \S9 provides a brief summary.
%Throughout this paper, we use a Concordance Cosmology with
%$\Omega_{\rm m}=0.25$, and $\Omega_{\Lambda}=0.75$
%and  parametrize the Hubble constant $\mathrm{H}_0$ via
%$h=\mathrm{H}_{0}/100$.
%When not explicitly specified otherwise, 
%absolute magnitudes are instead computed assuming
%$H_{0}=70~\mathrm{km~s^{-1}~Mpc^{-1}}$. 
\\
\section{Survey strategy and design}
\label{design}
\subsection{Star-galaxy separation and target definition}
VIPERS was conceived in 2007, focusing on the study of clustering and redshift-space distortions at
   $z\simeq 0.5-1.2$, 
   but with a desire to enable broader goals involving large-scale
   structure and galaxy evolution.
   The survey design was also strongly 
   driven by the specific features of the
   VIMOS spectrograph, which has a relatively small field of view 
   compared to fibre-fed instruments ($\simeq 18 \times 16\, {\rm arcmin}^2$), but a 
   larger yield in terms of redshifts per unit area (up to 1000
   spectra per exposure: \citealt{vvds_tech}).  

\begin{figure*}
\resizebox{\hsize}{!}{\includegraphics[clip=true]{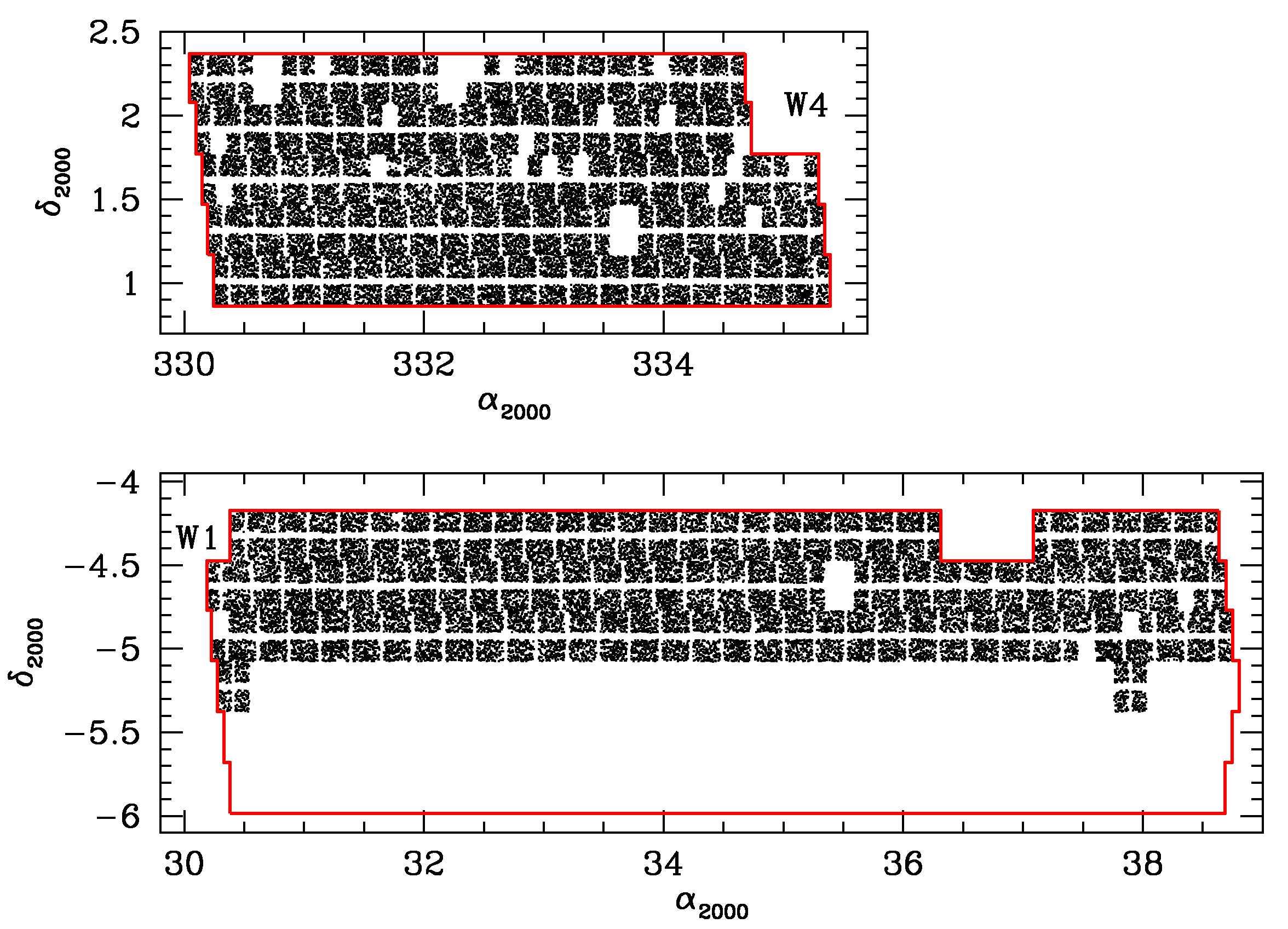} }
\caption{VIPERS survey areas. Black areas are the spectroscopically observed pointings. The thick red
line delimits the full area at the survey end. In the current release, some holes
due to bad observing conditions are present. Some of them will be filled before
the end of the survey.
}
\label{coverage}
\end{figure*}

The VIPERS overall sky coverage and field layout is shown in Figure
\ref{coverage}; the solid red line delimits the planned area, while the 
black dots show the spectroscopically observed objects. 
Table~\ref{surveyTable} shows the area planned in each of the two survey
fields as well as the area already spectroscopically covered, which is the subject 
of the current data release. The effective area is computed taking into account only those
portions of sky effectively exposed, i.e. not considering failed quadrants and the dead
cross between VIMOS quadrants.  
The last column in Table
~\ref{surveyTable} gives the area where photometry is reliable (see Section~\ref{masks}).
%data summary table
\begin{table*}
\caption{VIPERS survey field positions and area coverage: Surveyed area is the total area on sky, effective area is net of 
VIMOS cross and failed pointings/quadrants.}
\label{surveyTable}
\centering
\begin{tabular}{c c c c c c c }
\hline\hline
Field             & R.A$^1$   & Dec$^1$         & Final                  & PDR-1                     & PDR-1            &  PDR-1 area\\
                  &           &                  &surveyed  area        &  surveyed area           & effective area   &with good photometry\\
% observed objects are all entries in Wx_spectr_v3
\hline                                                                
$0226-04$ (W1) & 02h26m00.0s & $-04$deg$30'00''$ & $15.701\,{\rm deg}^2$ & $7.932\,{\rm deg}^2$    & $5.478\,{\rm deg}^2$ & $5.347\,{\rm deg}^2$   \\   
$2217+00$ (W4) & 22h17m50.4s & $+00$deg$24'00''$ & $ 7.851\,{\rm deg}^2$ & $7.851\,{\rm deg}^2$      & $5.120\,{\rm deg}^2$ & $4.968\,{\rm deg}^2$     \\    
\hline
\end{tabular}
\begin{tabular}{l c}
$^1$ Center of Field
&~~~~~~~~~~~~~~~~~~~~~~~~~~~~~~~~~~~~~~~~~~~~~~~~~~~~~~~~~~~~~~~~~~~~~~~~~~~~~~~~~~~~~~~~~~~~~~~~~~~~~~\\
\end{tabular}
\end{table*}
% macro coverage.mac
Given the luminosity function of galaxies and results
from previous VIMOS surveys (VVDS Deep and Wide: \citealt{vvds_main,vvds_wide}; 
zCOSMOS: \citealt{zcosmos_10k}), it was known that
a magnitude-limited sample with $i_{\rm AB}<22.5$ would cover
the redshift range out to $z\sim1.2$, and could be assembled with
fairly short VIMOS exposure times 
($<1$ hour).  Also, taking 2dFGRS as a local
reference, a survey volume around $5 \times 10^{7}\,h^{-3}{\rm Mpc}^3$ could be
explored by observing an area of $\simeq 24$~deg$^2$.  

Building upon this experience,
VIPERS was designed to maximize the number of galaxies observed in
the range of interest, i.e. at $z>0.5$, while at the same time
attempting to select against stars that represented a contamination
of up to 30\% in some of the VVDS-Wide fields (where by design no star-galaxy
separation whatsoever had been applied: see \citealt{vvds_wide}). Therefore we searched for
a criterion that could limit the stellar contamination without hampering
the completeness of the galaxy sample.

Stars and galaxies were separated using both 
their measured size  and their spectral energy distribution,
  derived through template fitting of the high-quality CFHTLS 5-band photometry. As
  shown by the extensive tests carried out using the fully sampled VVDS (Deep and
  Wide) data, the
  overall galaxy completeness and resulting stellar contamination can
  be optimized by applying the following criteria \citep{vipers_main}:
\begin{enumerate}
\item At $i_{\rm AB}<21$, stars are defined to be objects with $r_\mathrm{h}<\mu_\mathrm{rh} +
  3\sigma_\mathrm{rh} $. Galaxies are the complementary class. 
\item At $i_{\rm AB} \ge 21$, the previous criterion would exclude small compact galaxies from the target sample.
For these reason, objects are discarded as {\it bona fide} stars if 
  $r_h<\mu_{\mathrm{rh}} + 3\sigma_\mathrm{rh} $ AND
  $\log_{10}(\chi^2_\mathrm{star})<\log_{10}(\chi^2_\mathrm{gal})+1$,
\end{enumerate}
where $r_\mathrm{h}$ is the half-light radius of an object; $\mu_{\mathrm{rh}}$ and
$\sigma_{\mathrm{rh}}$ are the mean and standard deviation of the $r_\mathrm{h}$
distribution; while $\chi^2_\mathrm{star}$
 and $\chi^2_\mathrm{gal}$ result from the spectral energy distribution template fitting applied to the CFHTLS photometry.

The desired redshift range ($z>0.5$) for the objects classified as galaxies
is then selected requiring that 
\begin{equation}
(r-i) > 0.5 (u-g)  \;\;\;\ \lor \;\;\;   (r-i)>0.7 \, 
\label{col-col}
\end{equation}
(see Figure 3 in \citealt{vipers_main}).

\subsection{Additional AGN targets}
\label{AGNselection}
To broaden the scientific yield of the project, without affecting the
purity of the galaxy selection function, it was also decided to
supplement the target list  with a small number of AGN (Active Galactic Nuclei) candidates:
(1) a set of 449 X-ray selected candidates from 
the XMM-LSS survey in the W1 field \citep{xmm_lss}, supplied by
the XMM-LSS Consortium and added as compulsory targets (average of one object
every two quadrants); and (2) a set of
3696 photometrically defined AGN candidates, selected among stellar
objects resulting from the previous star-galaxy separation.  In a first version, an object was
classified as an AGN candidate if its colours satisfied the following criteria:

\begin{equation}
{\rm CC1:}
\left\{ \begin{array}{l}
(g-r)<1 \land (u-g)<0.6\\
(g-r)<1 \land 0.6 \le (u-g) < 1.2 \\
~~~~~~~~ \land (g-r)>0.5(u-g)+0.036\\
(g-r)<1 \land 0.6 \le (u-g) < 2.6 \\
~~~~~~~~ \land (g-r)<0.5(u-g)+0.214\\
(g-r)<1 \land (u-g)>2.6 \,.
\end{array} \right.
\label{CC1AGN}
\end{equation}

These criteria were calibrated on the VVDS point-like objects brighter than $i_{\rm AB}=21$,
and we expected a
sample with a completeness of $87.5$\% and
with a stellar contamination of $36$\%.  These percentages were not
confirmed by the actual data at the end of the first observing
season, indicating a significantly higher than expected contamination from stars.  
While investigating the origin of this discrepancy, we decided to revise criteria towards a more
restrictive definition, following which an AGN candidate must
satisfy  CC1 (equation \ref{CC1AGN}) and :
\begin{equation}
{\rm CC2:}
\left\{ \begin{array}{l}
(u-g)<0.6 \land -0.2<(g-i)<1\\
0.6\le (u-g)<1 \land -0.2<(g-i)<0.2\\
(u-g)\ge 1 \land (g-i)<0.6 \, .
\end{array} \right.
\label{CC2AGN}
\end{equation}
This more stringent definition has been applied very early in the survey (from 2010 onwards) so that
only 1\% of the spectroscopically confirmed AGN satisfy CC1 only.

AGN redshifts from photometrically defined candidates are included in the released
catalogue, but the precise selection function to
correct for any incompleteness is still being verified.
Brighter than $i_{\rm AB}=21$, where our 
selection for point-like AGN had been tested, the number counts of the AGN
candidates and the confirmation rate are as expected. 
Fainter than that,
where AGN have been observed as part of the primary target sample, the selection
function changes. A complete analysis of the sample of spectroscopically confirmed AGN 
will be the subject of a separate paper. The 189 XMM-LSS AGN candidates observed so far are not
included in PDR-1, and will be analysed separately.
%
%
%A first group of 242 correspond to
%about half of the 449 XMM-LSS X-ray mandatory candidates in W1 (Pierre et al., in
%preparation ???? ).  Note that a few of these are not confirmed as
%AGN.  Then, the colour-colour selected candidates turn out to have
%been affected in some unexpected way by the (previous) star-galaxy
%separation process.  We find a clear difference in the confirmation
%rate and number counts of the AGN candidates brighter than
%$I_{\rm AB}=21$, i.e. the threshold where the classification is
%purely geometrical.  In this case, the number counts of the AGN
%candidates and the confirmation rate are as expected.  
%
%
%
%HERE ARE THE FACTS:
%
%There are 2432 candidates with i<21, out of 3696 total colour-selected
%candidates%
%
%Of these 816 in total have been {\bf observed} (not CONFIRMED!).  531 of these have $i_{\rm AB}<21$.
%
%- At any magnitude: 478 out of 816 are confirmed as spectroscopic AGN
%
%- At i<21, 357 out 531 are confirmed as spectroscopic AGN  (success
%rate of 357/531 = 67.2\%)
%
%- This implies that the confirmation rate at mags fainter than
%$i_{\rm AB}=21$ drops to (478-357)/(816-531) = 121/285 = 42.5\%
%
%THESE ARE ONLY THOSE INITIALLY ADDED AS COLOUR-SELECTED AGN
%CANDIDATES (and we know have been affected by the a priori star-galaxy SED recovery of
%galaxies for $>21$).  THESE NUMBERS DO NOT INCLUDE THE AGN
%RE-DISCOVERED AMONG THE GALAXY TARGETS.
%
%- The total number of  $Z_{FLAG}=12,13,14$ is 930.   
%Of these, 474 are brighter than $i_{\rm AB}=21$.
\\
\section{VLT-VIMOS Observations}
\label{observations}
The VIPERS survey is being performed in the framework of the ESO Large programmes and is carried out using VIMOS on `Melipal', the 
Very Large
Telescope (VLT) Unit 3 \citep{vvds_tech}. VIMOS (Visible Imager Multi Object
Spectrograph)
makes use of slits cut out from masks. Its field of view of $\simeq 18 \times 16\, {\rm arcmin}^2$ is divided into four
quadrants, each having an area of $\simeq 7 \times 8.1\, {\rm arcmin}^2$. 
Each quadrant corresponds to an independent spectrograph and the four
spectrographs
observe in the same configuration for the same exposure time.

The standard VIMOS observing procedure requires the acquisition of a direct image, which
is used for mask preparation with the {\it vmmps}  software \citep{vmmps}
distributed by ESO: 
%a catalogue with the pixel position  of the ~60 brightest
%objects must be built by the user (in our case we used SEXtractor in automatic
%mode, \citet{sextractor} ); using a first guess for the Sky to CCD transformation, the sky coordinates of such
%objects are cross matched with the catalogue of the sources to be observed (user
%catalogue) and an accurate Sky to CCD transformation is computed by {\it vmmps}.
%This transformation, based on a $3\times3$ matrix, can accurately take into account
%image distortions, and has a typical rms of 0.25 arcsec. By applying it to the
%user catalogue, the pixel coordinates for each object are computed. Two {\it
%reference objects} per quadrant are manually chosen to allow for accurate
%acquisition of the field: a $6\times6$ arcsec square will be cut around each reference object.
%Such objects must be compact and bright, so that an accurate centroid can be
%computed during the short (typically a couple of minutes) acquisition image.
%Finally, within the user's catalogue {\it vmmps} automatically chooses the scientific objects to be
%observed, maximizing the number of slits to be cut in the mask, while at the
%same time taking into account the slit alignment and spectrum length
%constraints. During optimization, 
{\it vmmps}  assigns the slit length taking
into account object dimensions
and sky subtraction regions as specified by the
user. The minimum slit length can be changed by the software as part of the
optimization process, to maximize the number of slits per quadrant.
The files containing the mask definition are sent to ESO (via the P2PP tool) for
mask cutting before spectroscopic observations. 
%These are organized in Observation Blocks (OB). Each OB is
%composed of an acquisition procedure, a number of scientific exposures, and
%calibration exposures (dome flat fields and arc lamps). 
%The field acquisition entails the use of a direct image taken through the
%mask; the position of the reference objects is automatically computed and
%compared to that of the reference hole centres, and pointing shifts
%required to bring the objects to the centre of the reference holes are
%automatically computed and applied. Once the reference objects 
%are well centred in the reference holes, the grism is inserted and the
%spectroscopic exposures are carried out. 
As VIMOS suffers from some flexure
problems, calibration exposures are performed immediately after the scientific
exposures, maintaining the instrument at the same rotation angle as the scientific
exposure and inserting a screen at the Nasmyth focus. This ensures that we have
calibration lamps with the same flexure-induced distortions as the scientific
images, thus allowing for a more precise wavelength calibration. The instrument has no
atmospheric dispersion compensator; in order to avoid spectra distortions due to atmospheric refraction,
observations are confined within $\pm2$ hours from
the meridian.  

\begin{figure}
\resizebox{\hsize}{!}{\includegraphics[clip=true]{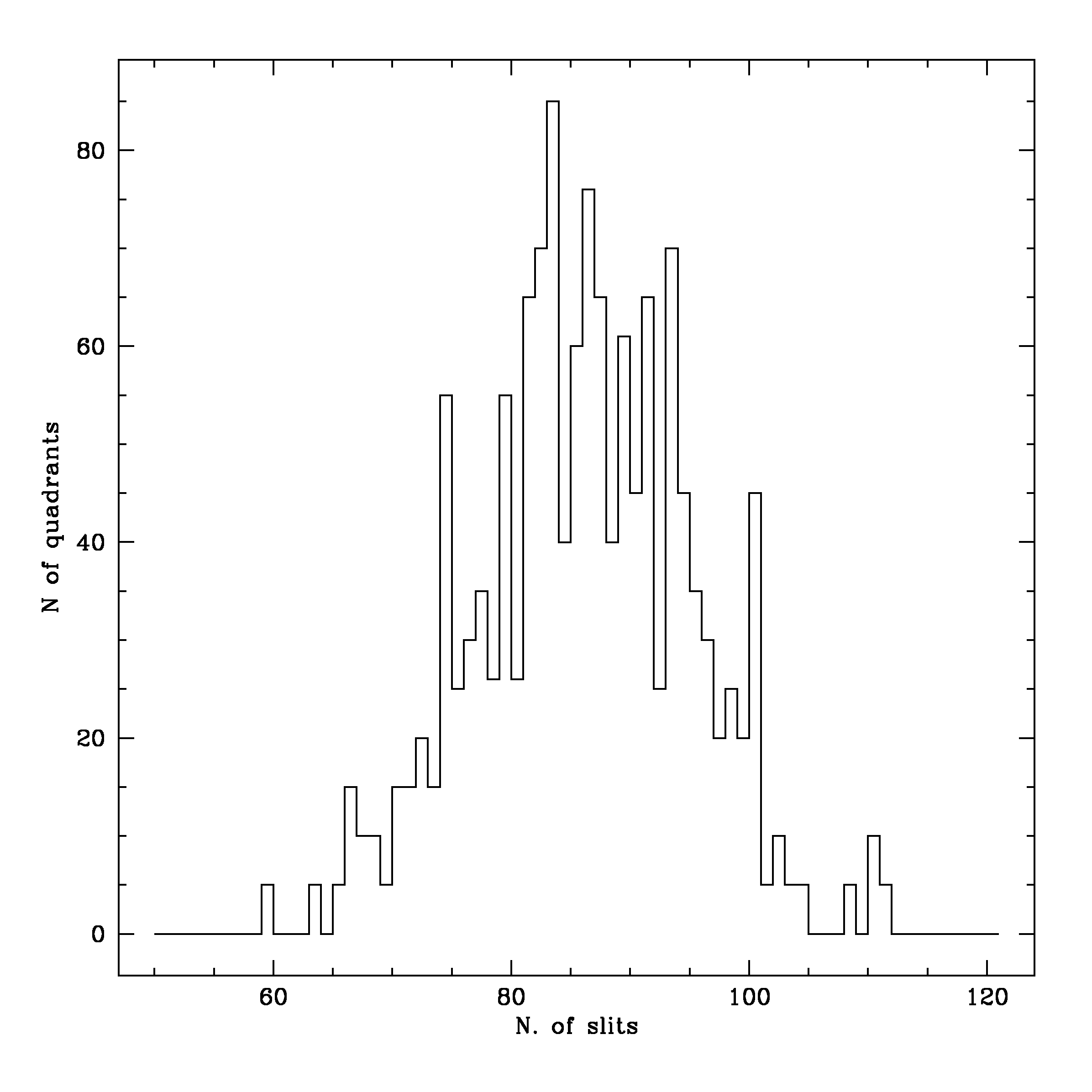}
}
\caption{The distribution of the number of slits per VIPERS quadrant.}
\label{slitNumber}
\end{figure}

Given the chosen magnitude limit ($i_{\rm AB} \le 22.5$), the total exposure time 
adopted for the VIPERS spectroscopic observations is 2700 sec. The observation of  one pointing is split 
into five exposures of nine minutes each. Observations have been carried out aligning slits along the East-West direction, so that atmospheric refraction effects are minimized.
We use 1-arcsec wide slits, a value which well matches
the average seeing in Paranal. The
`Low-Resolution Red' (LR-Red) grism is used, providing a spectral
resolution $R\simeq 230$ and a mean dispersion of 7.3\AA/pixel.  
When preparing masks,
we have allowed for a minimum of 1.8 arcsec per side for sky subtraction.
In order 
to further boost the survey sampling and the multiplexing capabilities of VIMOS, 
we have followed the approach 
described in \citet{short_slits}: essentially, we have arbitrarily assigned  
an object size of 0.5 arcsec to all VIPERS sources. In this way, {\it vmmps}
has more freedom in the slit positioning optimization process, and we have been
able to increase the number of objects per quadrant by $\sim 10\%$, reaching a median of 87 slits per quadrant
(see Figure \ref{slitNumber}).
The drawback of this approach is that bigger objects have a non-negligible chance of filling most of the slit, 
and sometimes the automatic spectral extraction may fail. 
{\it A posteriori} we have verified  that
the median slit length is 7 arcsec, and that only a few among the brightest objects suffer from this problem.  

% slitNumber distribution macro slitNumber.mac

\section{Spectroscopic observations and data reduction} 
\label{reduction}
\subsection{Quality of VIMOS raw data}
\begin{figure}
\resizebox{\hsize}{!}{\includegraphics[clip=true]{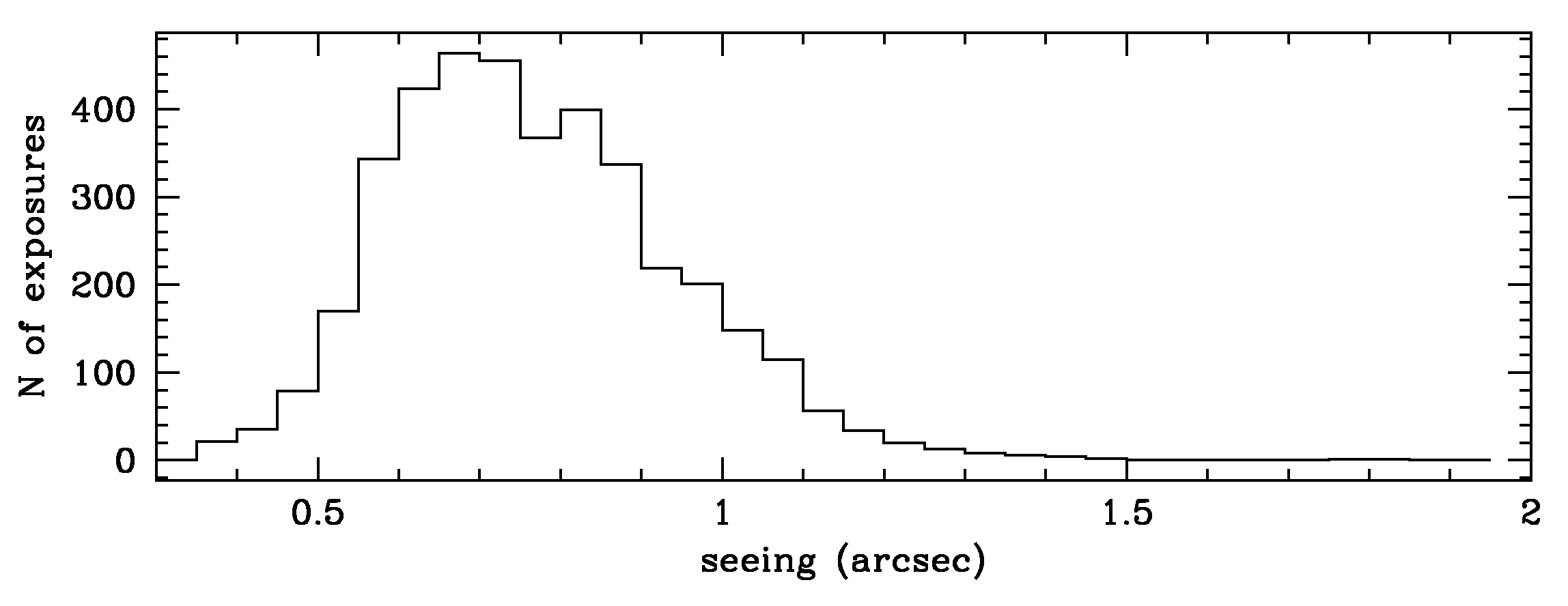}
}
\caption{The distribution of seeing values for VIPERS observations.}
\label{seeing}
\end{figure}

VIPERS observations started in September 2008 and have proceeded at a steady rate since then. The
median airmass of PDR-1 data is 1.14, and remains below 1.3 in 90\% of the
exposures. To match the slit width and minimize slit losses, the requested
maximum seeing was 1 arcsec. 
Figure~\ref{seeing} shows the  seeing distribution
as measured from the scientific exposures. 
The median seeing is 0.8 arcsec and 90\% of exposures have a seeing below 1.05 arcsec. 
85\% of observations have been carried out in
dark time, while the remaining fraction either had a moon illumination below 20\%, or
a moon distance above 70 deg.
\\

% seeing distribution macro seeing.mac
In spring/summer 2010, VIMOS was upgraded with new red-sensitive CCDs in each of
the 4 channels
\citep{vimos_upgrade}. The original thinned E2V detectors
were replaced by twice as thick E2V devices, considerably lowering
the fringing and increasing
the quantum efficiency by up to a factor two over the wavelength
range of the LR-Red grism. 
The lesser (almost non-existent) fringing allows for  a better sky
subtraction redwards of 7500\AA ~(where OH bands dominate the sky emission),
which improves the data quality. 
In the following, we use the term `epoch 1' to denote data acquired before
the refurbishment, and `epoch 2' for data acquired afterwards. 
Overall, 22049 targets have been observed in epoch 1, 
and 40813 in epoch 2 (see Table \ref{detectionTable}).
\begin{figure*}
\resizebox{\hsize}{!}{\includegraphics[clip=true]{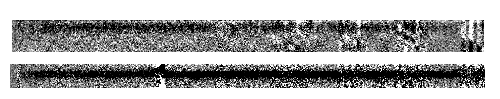}}
\caption{A typical 2D extracted spectrum from
epoch 1 data (top) and epoch 2 data (bottom). Wavelength ranges 
from 5600\AA ~to 9500\AA ~from left to right.
Spatial direction is along the {\it y} axis.}
\label{noise2D}
\end{figure*}
In
Figure~\ref{noise2D} we show a typical 2D extracted spectrum from
epoch 1 data (top) and epoch 2 data (bottom). The improved sky subtraction in
epoch 2 data is clearly visible. More quantitatively, Figure~\ref{noisedist} shows
the distribution of the r.m.s. of the sky subtraction residuals in the 7500-9100\AA ~wavelength range for  epoch 1
(empty histogram) and epoch 2 (shaded histogram) data. The median of the distribution diminishes by a factor $2.6$,
clearly demonstrating the significant improvement in the sky subtraction.
In Section~\ref{spectroSample} we will show how this affects the redshift
measurement.
\begin{figure}
\resizebox{\hsize}{!}{\includegraphics[clip=true]{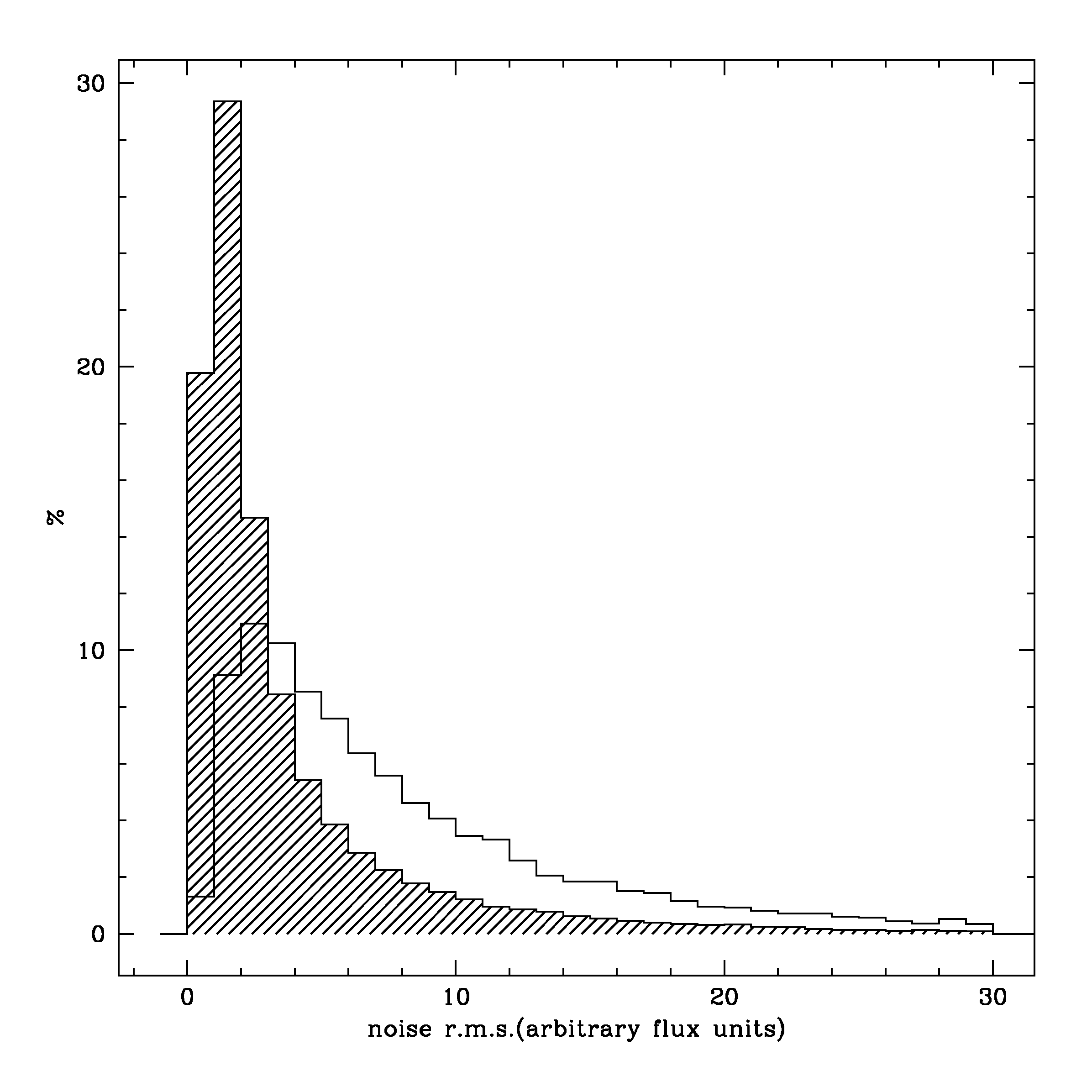}
}
\caption{Median noise r.m.s. distribution of the VIPERS spectra in the 7500-9100\AA~wavelength range. 
Empty histogram refers to data acquired before VIMOS refurbishment, 
shaded histogram to those acquired after the refurbishment.}
\label{noisedist}
\end{figure}
%
%epoch1 median noise (quartile 1, median, quartile3) 5.8 8.3 13.
%epoch2 median noise (quartile 1, median, quartile3) 3.6 5.1 8.3
%
%epoch1 median errnoise (quartile 1, median, quartile3) 3.3 6.3 12
%epoch2 median errnoise (quartile 1, median, quartile3) 1.1 2 4.4
The VIMOS upgrade also included substituting the old passive flexure compensation 
system with a new active flexure compensator (AFC), which should be
more reliable
and precise. Unfortunately, for a few months the AFC was operational only during
target acquisition. In other words, 
some observing runs were actually carried out without any flexure compensation system during the
spectroscopic exposure. This implies that during the same OB 
the images of the slits moved up to 2 pixels on the detector plane. To account for this effect,
we have modified the reduction pipeline to perform the slit tracing exposure by exposure. 
Since the arc lamp is always 
acquired at the end of the OB, its tracing is slightly different from the 
tracing of the first exposure, implying that the wavelength calibration at 
the slit edges is not as good as at the slit centre. This has a minor effect on the quality 
of the sky subtraction: unless
the object is very far from the slit centre, the only consequence is that 2D spectra are noisier close to the edges.
In Section~\ref{spectroSample} we will show that the redshift measurement is statistically unaffected.

\subsection{Data reduction procedure}
VIPERS  data reduction
is performed with a fully automated pipeline, starting from the raw
data and flowing down to the wavelength- and flux-calibrated spectra and redshift measurement. 
The pipeline
is an updated version of the algorithms and dataflow from the original VIPGI system, 
fully described 
in \citet{vipgi}. We summarise here the main concepts.

As a first step, in each raw frame the 2D dispersed 
spectra are located and 
traced. Each raw spectrum is collapsed along the dispersion direction, and the
object location computed.  A first sky subtraction is performed row by row,
avoiding the region identified as the object.
An inverse dispersion solution is computed for each column of each dispersed
spectrum making use of an arc calibration lamp. The wavelength calibration
residuals distribution is well peaked around 0.771\AA ~(i.e. 1/10 of a pixel) and
97\% of spectra have a wavelength calibration uncertainty below 1.25\AA ~(1/5 of a pixel).
The inverse dispersion solution is applied before extraction.
A further check on the wavelength of sky lines is computed  on the linearized 2D
spectra, and, if needed, a rigid offset applied to data in order to bring the sky
lines to their correct wavelength. The different scientific exposures of the same
field are registered and co-added, and a second background subtraction is
performed repeating the procedure carried out before. Finally, 1D spectra are
extracted applying the Horne extraction algorithm \citep{horne}, and spectra are 
corrected for the instrument sensitivity function, as derived from the standard
spectrophotometric observations routinely carried out by ESO.
Given a 2D spectrum, 
we can compute sky subtraction residuals as
\begin{equation}    
\label{residual}    
    \langle{\rm skyResidual}\rangle_{i}= \frac{\sum_{j \in {\rm NRegion}} (\langle C\rangle-C_j)^2}{{\rm NRegion}-1} ,
\end{equation}        
where NRegion is the number of pixels in the sky region on one side of the object,  
$\langle C\rangle$ is the mean of the counts in the sky region and  $C_j$ are the counts in pixel $j$, 
at wavelength $i$.
This computation is done on both sides of the object (resL and resR), 
taking into account slit borders and possible second objects in the slit.
From such residuals, a mean noise spectrum is computed, 
the noise in the $i$-th pixel being given by
\begin{equation}        
\label{noiseSpectrum}    
   {\rm noise}_{i}^2=\frac{{\rm resL}_i+{\rm resR}_i}{2}+S_i=\langle{\rm skyResidual}\rangle_i + S_i ,
\end{equation}        
where $S_i$ are the source counts at wavelength $i$.
To obtain fully flux calibrated spectra, while correcting for slit losses, we
have convolved each spectrum with the
CFHTLS {\it i} filter response function, and computed a normalization factor between
the spectrum and the photometric magnitude.
We then run EZ \citep{EZ} on all 1D extracted spectra for a first automatic redshift
measurement.
All data reduction has been centralised in our data
reduction and management centre at INAF -- IASF Milano. When ready, the fully
reduced data are made available to the team within a dedicated
database.  The full management of these operations within the
{\it EasyLife} environment is described in \citet{easylife}.

\subsection{Redshift estimation, reliability flags and confidence levels}
\label{redshift}
In most cases, the final redshift measurement is performed and validated by two
 team members independently.
The reliability of the measured redshifts is quantified at the time of
validation following a scheme similar to that used for the VVDS \citep{vvds_main} and
zCosmos surveys \citep{zcosmos_main}.  Measurements of stars and galaxies
are flagged using the following convention: 

\begin{itemize} 
\item
    Flag 4: a highly reliable redshift (estimated to have $>95\%$
    probability of being correct), based on a high SNR
spectrum and supported by obvious and consistent spectral features.  
\item
    Flag 3: also a very reliable redshift, comparable in confidence
    with Flag 4, supported by clear spectral features in the spectrum, but not
necessarily with high SNR. 
\item
    Flag 2: a fairly reliable redshift measurement, but not as
      straightforward to confirm as for Flags 3 and 4,  supported by
    cross-correlation results, continuum shape and some spectral
    features, with expected chance of $\simeq 75\%$ to be
    correct. We shall see in the following that the actual estimated confidence
  level will turn out to be significantly better.  
\item
    Flag 1: a reasonable redshift measurement, based on weak spectral features
    and/or continuum shape, for which there is roughly a $50\%$ chance that
    the redshift is actually wrong.
\item
    Flag 0: no reliable spectroscopic redshift measurement was possible.
\item
    Flag 9: a redshift based on only one single clear spectral emission
feature.
\item
  Flag -10: spectrum with clear problems in the observation or data
  processing phases. It can be a failure in the {\it vmmps} Sky to CCD
  conversion (especially at field corners), or a failed extraction by VIPGI
  \citep[][]{vipgi},  or a bad sky subtraction because the object is too
  close to the edge of the slit. 
\end{itemize}

Broad-Line AGN can be easily identified during the validation process from the width
of their emission lines. The flagging system for AGN is similar, though not
identical, to the one adopted for stars and galaxies:
\begin{itemize} 
\item
    Flag 14: secure AGN with a $>95\%$ reliable redshift, including at least
    2 broad lines; 
\item
    Flag 13: secure AGN with good confidence redshift, based on
    one broad line and some faint additional feature; 
\item
    Flag 19: secure AGN with one single secure emission line feature,
    redshift based on this line only;
\item
    Flag 12: a $>95\%$ reliable redshift measurement, but lines are not
    significantly broad, might not be an AGN;
\item
    Flag 11: a tentative redshift measurement, with spectral features not
    significantly broad. 
\end{itemize}

Serendipitous (also called secondary) objects appearing by chance within the slit of the main
target  are identified by adding a `2' in front of the main flag. In Section~\ref{sec:duplicates}
we will assess the actual confidence levels of the flags by comparing with results of other surveys.

Once  the final review of the
redshifts is complete, a decimal part of the flag `.X' indicating concordance
or discordance with the photometric redshift is added to the main flag. An automatic algorithm
cross-correlates the spectroscopic measurement ($z_{\rm spec}$) with the
corresponding photometric redshift ($z_{\rm phot}$), estimated from the
five-band CFHTLS photometry using the {\it Le Phare} code
\citep{vvds_photoz,lephare}.  The 68\% confidence interval $[z_{\rm phot-min},
z_{\rm phot-max}]$ (in general not symmetric)  based
on the PDF of the estimated $z_{\rm phot}$ is provided by {\it Le Phare}.  If 
$z_{\rm spec}$ is included within the 
$z_{\rm phot-min}-z_{\rm phot-max}$ range, 
spectroscopic and photometric redshifts  are considered in agreement and a
flag 0.5 is added to the primary flag.  
Redshifts are instead in marginal agreement (and a flag 0.4 is added)
when they are comparable only at the $2\sigma$ level, where $2\sigma$ is 
the minimum (maximum) between 
$z_{\rm phot}-0.05\times(1+z_{\rm phot})$ and $z_{\rm phot-min}$ ($z_{\rm phot}+0.05\times(1+z_{\rm phot})$ and $z_{\rm phot-max}$), 
being 0.05 twice the median scatter of the comparison between $z_{spec}$ and $z_{phot}$.
This allows us to signal cases in which the PDF of the single measurement is  rather narrow, 
but still the spectroscopic redshift is close. Finally, we
add `0.2' to the redshift flag when neither of the two criteria is
satisfied, and `0.1' when no $z_{\rm phot}$ estimate is available.

Thus, whatever the primary integer flag is, a flag `*.5' or '*.4' is an
indication supporting the
correctness of the redshift. This is particularly useful in the case
of highly uncertain, ${\rm flag}=1$ objects, for which the confidence level can be
increased by the agreement with the photometric redshift value. 
In all VIPERS papers redshifts with flags ranging between 2.X and
9.X %(or 12 and 19 in the case of AGN)
are referred to as reliable redshifts and are the only ones normally
used in the science analysis.

% topcat table NEWEST_all_clean.fits
% subset secure density redblue log full subrange
% log axis fonts latex 19 bold function x+/-0.15*x

\begin{figure}
\resizebox{\hsize}{!}{\includegraphics[clip=true]{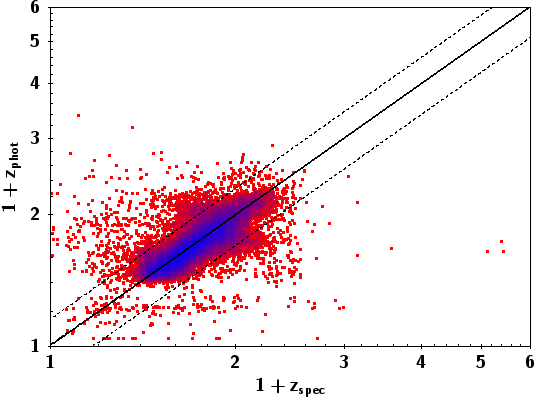}}
\caption{Comparison between photometric and spectroscopic secure redshifts in the VIPERS survey. The dashed lines
are for $z_{phot}=z_{spec}\pm 0.15(1+z_{s})$}
\label{photoz}
\end{figure}

In Figure~\ref{photoz} we show the comparison between photometric 
and spectroscopic 
redshifts  for the subsample of secure spectroscopic redshifts. The dispersion,  
computed as the normalised median absolute deviation
$\sigma_{\Delta z/(1+z_{s})}=1.48*median(|\Delta z|/(1+z_{s}))$, 
is 0.03 while
the outlier rate, i.e. the proportion of objects with
$|\Delta z|\geq (1+z_{s})$, is 5.12\%, in excellent agreement with what obtained
by \citet{CFHTLS_photoz} for the CFHTLS T0004 fields.

\subsection{Repeated observations and redshift errors}
\label{sec:error}
The footprint of VIMOS is not a perfect rectangle, as each quadrant has a
slightly different width. When designing the survey pointings, we have 
allowed for a small overlap (of the order of few arcseconds) between
adjacent pointings.
For this reason a number of VIPERS targets have been observed twice. 
Objects falling within this overlapping region
have some chance of being observed twice, and this happened
for 783 objects in the VIPERS survey. Additionally, during the
re-commissioning of VIMOS after the CCD refurbishment in summer 2010,
a few pointings were re-observed to verify the performance with the
new setup \citep{vimos_upgrade}, targeting further 1357
objects. 
In total, this results in a very useful
sample of 2143 objects observed at least twice. After excluding second objects, stars and AGNs, 
we are left with 1235 target  galaxies
yielding a reliable redshift (i.e. with a flag $\ge 2$) in both
measurements, of which 1192 have compatible measurement, i.e. the two redshifts differ by less than 
$\Delta/(1+z) < 3 \sigma_z \simeq 0.0025$. This subsample
can be used to obtain an estimate of the
internal r.m.s. value of the redshift error of VIPERS galaxies, as
discussed in \citet{vipers_main} and reported here for completeness.
% macro dup.mac in ../perIlPaper/duplicates
\begin{figure}
\resizebox{\hsize}{!}{\includegraphics[clip=true]{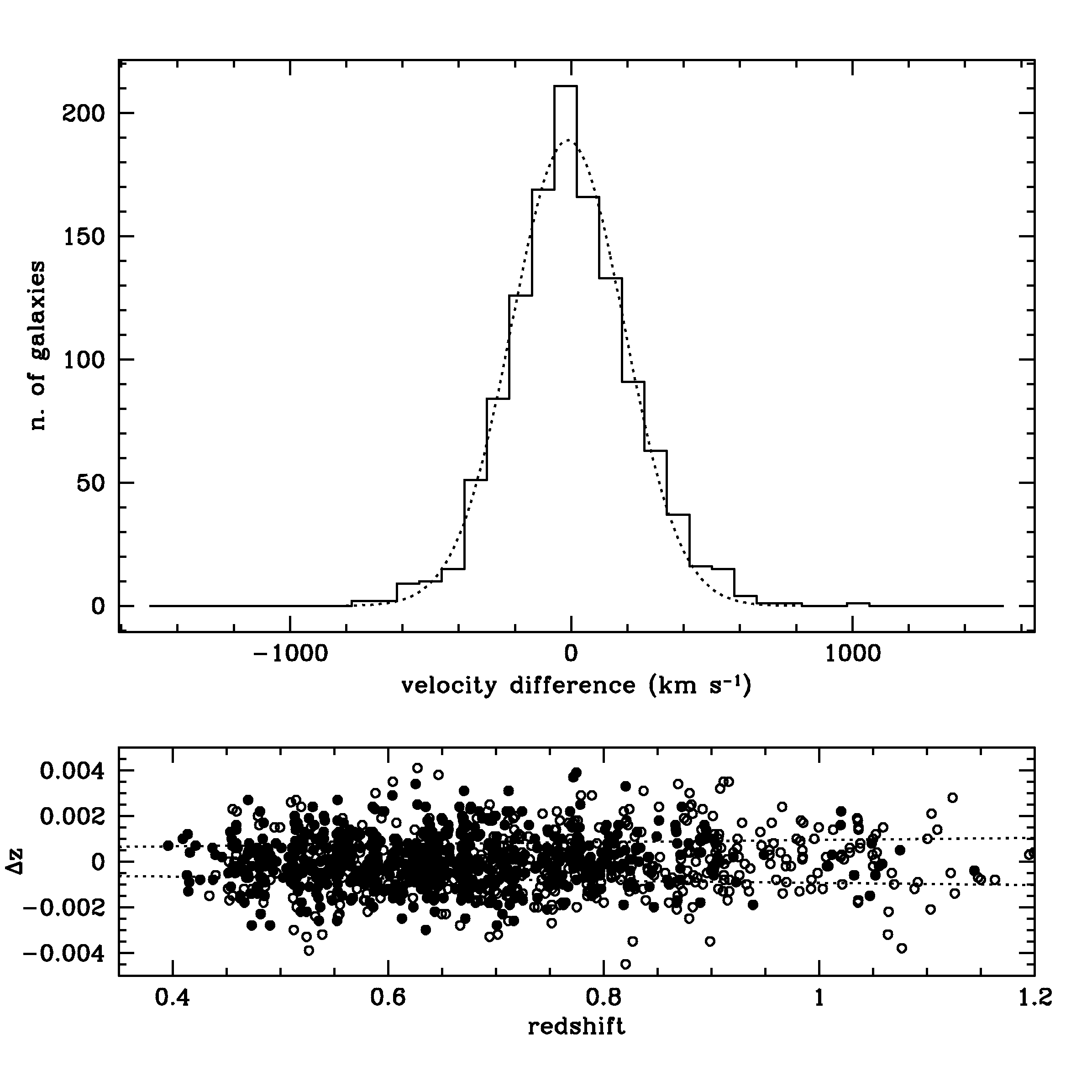}}
\caption{
The distribution of the differences between two
       independent redshift measurements of the same object, obtained
       from a set of 1192 VIPERS galaxies with redshift flag $\ge 2$.
Top: distribution of the velocity 
        differences $\Delta v = c\Delta z /(1+z)$. Catastrophic failures, 
defined as being discrepant by
       more than $\Delta z = 6.6\times 10^{-3} (1+z)$, have been excluded. 
The best-fitting Gaussian has a dispersion of $\sigma_2=200$ km s$^{-1}$,
       corresponding to a single-object {\it rms} error
       $\sigma_v=\sigma_2/\sqrt{2}=141$  km s$^{-1}$. In terms of
       redshift, this translates into a standard deviation of
       $\sigma_z=0.00047(1+z)$ for a single galaxy measurement.
       In the bottom panel, the darker dots correspond to 
       highly reliable redshifts (i.e. flags 3 and 4), which show a
       dispersion substantially similar to the complete sample (see
       text).  
} 
\label{double-obs}
\end{figure}
In Figure~\ref{double-obs}, top panel, we show the distribution of rest frame velocity difference 
for these compatible measurements. The distribution is well fitted by a Gaussian centred on the origin 
with a standard deviation of $\sigma= 200\,{\rm km\,s}^{-1}$,
corresponding to a single-object $1\sigma$
error $\sigma_v=\sigma_2/{2}^{1/2}=141\,{\rm km\,s}^{-1}$. In terms of
redshift, this yields a standard deviation on the redshift
measurements of $0.00047(1+z)$.  
Using only the  most reliable 
spectra (i.e. flags 3 or 4 in both measurements), we are left with 648
double measurements and the resulting rest-frame 2-object dispersion
decreases to $\sigma_2=195$ km s$^{-1}$. This
indicates that flags 9, 2, 3 and 4 are substantially equivalent in terms
of redshift precision when the redshift is correct. In the next section we will discuss 
the confidence level of redshifts for each flag. 
\\
%comparison table
\begin{table*}
\caption{Confidence of redshift flags  }
\label{zcompaTable}
\centering
\begin{tabular}{c c c c c}
\hline\hline
VIPERS Class & Comparison Class &N. of  Common Galaxies  &
Concordant Redshifts &  VIPERS Confidence Level \\
(Flags)   & (Flags)  & &  & \\
\hline
\multicolumn{5}{c}{VIPERS Internal}\\
\hline
3,4   & 3,4  & 654     &  648         &  99.5\% \\
2,9   & 3,4  & 385     &  373         &  98.4\%\\
2     & 2    & 151     &  129          &  92.4\%\\
\hline
\multicolumn{5}{c}{VIPERS vs. VVDS}\\
\hline
3,4  & 3,4     & 346     &  336          &  98.5\% \\
2,9  & 3,4     &  74     &   66           &  90.6\%\\
%2,9  & 2,9,3,4 & 255     &  220         &  92.9\%\\
\hline
\multicolumn{5}{c}{VIPERS vs. PRIMUS}\\
\hline

3,4  & 3,4     & 1619    & 1478         &  95.5\%   \\
2,9  & 3,4     & 876     &  708         &  84.6\%\\
\hline\hline

\end{tabular}
\end{table*}
\\

%\subsection{Redshift comparison with external data}
\subsection{Statistical confidence levels corresponding to redshift flags}
\label{sec:duplicates}
A first estimate of the actual, statistical confidence level of the redshift flags
associated to our redshift measurements has been presented
in~\citet{vipers_main} using the available duplicated observations 
mentioned above. These results are reported in the top part of
Table~\ref{zcompaTable}. 
Here we extend and complement this analysis, making use of external
measurements. 

The confidence level of a given Flag class (e.g. Flag 3+4)  from VIPERS, which we call survey $X$ for simplicity,
can be estimated from the fraction of measurements in agreement among
those galaxies observed independently by another survey, $Y$, with comparable or
better reliability.  We note that while this latter aspect is easy to
evaluate when comparing VIPERS to 
a survey like VVDS, which was observed with a very similar set-up,
evaluating the intrinsic redshift reliability of a very different data
set, for which additionally we may only have literature information,
is less straightforward. This will be the case for the second comparison
case discussed below, i.e. the PRIMUS survey.

In practice,  if $N_{\rm meas}$ is the number of
common measurements and $N_{\rm agree}$ is the number of redshifts in
agreement\footnote{As above, two redshift 
measurements are defined to be `in agreement' when their 
difference $\Delta/(1+z) < 3\sigma _z \simeq 0.0025$.}, then the
probability for two redshifts to agree corresponds to the combined
probability that $X$ and $Y$ give the correct redshift, i.e.
\begin{equation}
P(X|Y) = P(X) \times P(Y) = \frac{N_{\rm agree}}{N_{\rm meas}} \,\,\,\, .
\end{equation}
If we are considering the same Flag category for both
data sets and assume that the flags indicate similar
confidence levels for the measurements, then $P(X)\simeq P(Y)$ and
thus  
\begin{equation}
P(X) = \sqrt{{N_{\rm agree}}\over{N_{\rm meas}}} \,\,\,\, .
\end{equation}
If on the other hand we are comparing two classes with different
intrinsic confidence levels, we need to estimate that of the
reference class (e.g. $P(Y$), to be able to compute $P(X)$, i.e. 
\begin{equation}
\label{eq8}
P(X) = {1\over{P(Y)}}{{N_{\rm agree}}\over{N_{\rm meas}}} \,\,\,\, .
\end{equation}
While for internal data it is often straightforward to estimate
$P(Y)$, this is non-trivial for data obtained from the literature,
unless it is certain that the reference class is of much higher reliability than 
the one for which we want to estimate the confidence. In such a case,
$P(Y)\simeq 1$ can be a reasonable assumption. 
\vskip 0.3truecm

\noindent {\bf Comparison to VVDS.} Some areas within both W1 and W4 fields have been observed
within the VVDS survey and a similar 
comparison between VIPERS and VVDS has been presented in
\citet{vvds_last}.
Let us make the reasonable assumption that the typical redshift error in VVDS and VIPERS are
comparable, with rms dispersion $\sigma_z$. We are aware that an
improvement in SNR has certainly been introduced for the VIPERS data
by the refurbishment  of the detectors in 2010; our assumption will
only make the comparison more conservative for VIPERS.  
There are 420 objects with a reliable (flag 2 through 9) redshift in  
VIPERS which have a highly reliable (flag 3 and 4) in VVDS. 

The results are shown in the central part of Table~\ref{zcompaTable}.
We first compared VIPERS Flag 3 and 4 redshifts with the 
same class of VVDS, obtaining a confidence level of 98.5\%; this is
very consistent with the value 99.6\% estimated internally using VIPERS
repeated observations.  
For the flags 2 and 9 together we instead
obtain a confidence level of 90.6\%, against the higher confidence,
98.4\%, estimated from the repeated VIPERS observations.

Conversely, if we assume the P(X) obtained from the VIPERS internal comparison,
%(First line of table 2) 
using equation~\ref{eq8} we can obtain an estimate of the VVDS flag 3,4 confidence level. 
The resulting confidence level for VVDS flags 3,4 is ~98\%.  
\vskip 0.3truecm

\noindent {\bf Comparison to PRIMUS.} W1 is also one of the areas
selected by the PRIMUS survey \citep{primus2}. 
Using only the PRIMUS reliable objects (PRIMUS flag 3 and 4), there are 2495 objects 
for which VIPERS measures a reliable redshift (flag 2 through 9) in VIPERS.
For this comparison, we consider two measurements to be concordant when
$\Delta/(1+z) < 3(\sigma_{z\rm VIPERS}^2 + \sigma_{z\rm PRIMUS}^2)^{1/2}$, where 
$\sigma_{z\rm PRIMUS}=0.003$ for PRIMUS flag=4 and
$\sigma_{z\rm PRIMUS}=0.015$ for PRIMUS flag=3 \citep{primus2} 

The results are shown in the bottom part of Table~\ref{zcompaTable}.
Both redshift classes considered (Flags 3+4 and 2+9) give slightly
smaller confidence level, when compared to PRIMUS.  The 
problem in interpreting these results is whether the confidence levels corresponding to the PRIMUS
flag system are similar to those of VIPERS or VVDS.  We note that PRIMUS spectra
have a coarser resolution than VIPERS ones, with $R$ degrading from
100 to 20 from blue to red.  Taking this into account, the agreement
of the Flag 3+4 classes for the two data sets is very good. In
particular, the difference in the results in the first lines (Flags
3,4 vs 3,4) for the `VIPERS Internal' and `VIPERS 
vs. PRIMUS' cases in the Table is probably more a test for the PRIMUS
best redshifts, rather than for VIPERS.  

As we did for the VVDS case, we can obtain an estimate of the PRIMUS flag 3,4 confidence level
assuming that VIPERS flags confidence level is 99.5\% (as obtained from the internal comparison),
The confidence level of 91.75\% we obtain is in very good agreement with 
the value quoted by  \citet{primus2}. 

Overall, the consistency of the estimates
performed internally (top) and with VVDS (middle), indicates that 
redshifts for Flags 3+4 in VIPERS should have a  99\% confidence of being correct,
while class 2+9 is expected to have a confidence level of at least 90\%,
probably close to 95\%.  If we compare these confidence levels to
those indicated in the original prescriptions for the flag
assignments, summarized in \S~\ref{redshift}, we see that in
particular for the Flag 2 objects the actual redshifts are more
reliable than expected.

\section{The VIPERS PDR-1 sample}
\label{spectroSample}
We define the observed target sampling rate as the ratio between 
the observed target objects and
all possible targets within the spectroscopic area. The detection rate is 
the percentage of detected targets over observed targets , while the redshift measurement success rate and reliable
measurement success rate
are the fraction of measured and reliably measured objects over the detected ones.
The X-ray AGN candidates are excluded from this computation, as they
are put in masks as compulsory objects (see Section \ref{AGNselection}), 
and as such follow a different selection function. In Table \ref{detectionTable} we show some basic statistics of the VIPERS PDR-1 spectroscopic sample, 
split for W1 and W4, and for epoch 1 and epoch 2 data. Numbers in parenthesis indicate the fractions of the
different spectroscopic flags.
So far, we have observed 63\,942 objects (62\,862 targets plus 1080 serendipitous objects) with a detection rate of 96\%. 
The detection rate shows no difference between W1 and W4, nor between epoch 1 and epoch 2 data (see section \ref{observations}).
The fraction of measured redshift, as well as  of reliable redshift is satistically higher for epoch 2 data, confirming the better quality of the 
data acquired after VIMOS refurbishing.
Table \ref{detectionTable} reveals the excellent quality of VIPERS spectroscopic data: globally the 
redshift measurement success rate is 94\%. 
81\% of the objects have a reliable redshift,
with 58\% of the redshift with a confidence level higher than 99\% (flag 3 and
4). In the last two rows in Table \ref{detectionTable} we show the detection rate and redshift measurement rate split into
epoch2a, when no flexure compensation was active during exposures, and epoch2b, when the active 
flexure compensator was working correctly. 
The identical detection  and measurement rates demonstrate that our reduction pipeline has successfully recovered
for this effect.    

% perc are det/obser, meas/det and reliable/det
\begin{table*}
\caption{VIPERS detection rate. Numbers in parenthesis indicate the fraction over the observed targets (column 3) and detected targets
(columns 4 and 5).}
\label{detectionTable}
\centering
\begin{tabular}{c c c c c }
\hline\hline
Field & Observed  & Detected   &  Measured  & Reliable \\
      &  targets &  targets  &   redshifts & redshifts \\
\hline     
W1 &        31602         & 30244  (96\%)    &       28376  (94\%)        & 24553 (81\%)           \\
W4 &        31260         & 29897  (96\%)    &       28041  (94\%)        & 24050 (80\%)           \\
\hline     
epoch 1 &    22049        & 21212  (96\%)    &       19209 (91\%)         & 15541 (73\%) \\
epoch 2 &    40813        & 38929  (95\%)    &       37208 (96\%)         & 33062 (85\%) \\
\hline     
epoch 2a &   10236        &  \phantom{0}9813 (96\%)    &   \phantom{0}9464 (96\%)         & \phantom{0}8495 (87\%) \\
epoch 2b &   30577        &  29116 (95\%)    &      27744  (95\%)         & 24567 (84\%) \\
\hline           
total &     62862         & 60141  (96\%)    &        56417 (94\%)      &   48603   (81\%)      \\
\end{tabular}
\end{table*}

\begin{figure}
\resizebox{\hsize}{!}{\includegraphics[clip=true]{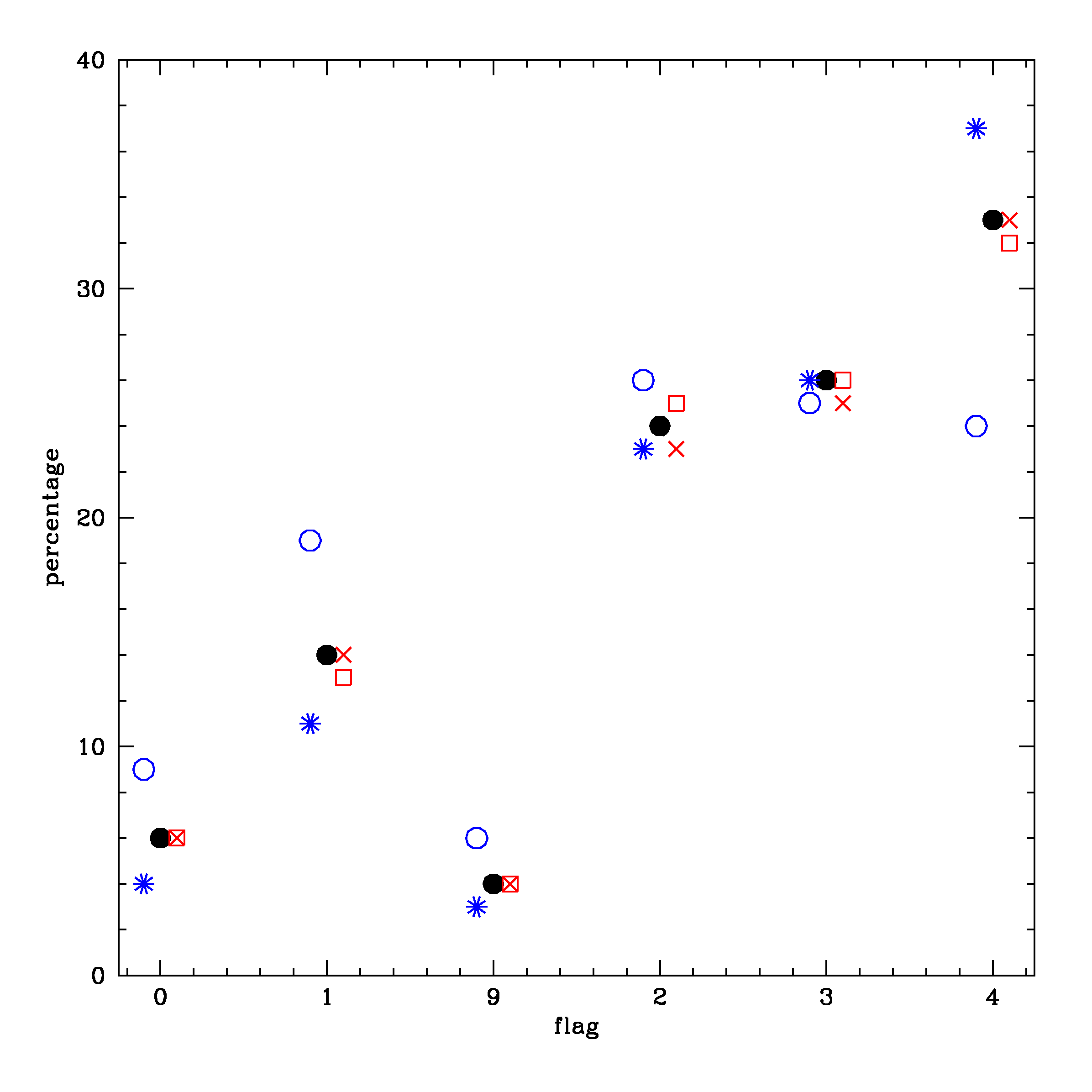} }
\caption{Fraction of redshift flags as from Table \ref{flagTable}:  filled points for the full sample, open red squares 
for the W1 sample, red crosses for the W4 sample, 
blue open circles and stars for epoch1 and epoch2 samples respectively. W1 and W4 points as well as 
epoch1 and epoch2 points have been shifted along the x axis for clarity.}
\label{percFlag}
\end{figure}

Table \ref{flagTable} and Figure \ref{percFlag} show the percentage of the various redshift flags. 
For the purpose of this table and figure, the distinction between AGN and galaxies/stars has been neglected
so that flags 1.x include also flags 11.x, flags 2.x include flags 12.x, and so on.
The better quality of epoch 2 data is clearly demonstrated by the much lower fraction of flags 0 (which decreases from 9\% to 4%)
and 1 (decreasing from 19\% to 11\%), and the much higher fraction of flags  4 (rising from 24\% to 37\%).

% redshift FLAGS percentages are wrt to detected targets above
\begin{table*}
\caption{VIPERS redshift flag distribution. Numbers in parenthesis indicate the fraction of the
different spectroscopic flags over detected targets.}
\label{flagTable}
\centering
\begin{tabular}{c c c c c c c }
\hline\hline
Field &  Flag 0.x  & Flag 1.x & flag 9.x & flag 2.x & flag 3.x & flag 4.x  \\
\hline           
W1     & 1863  (6\%)   & 3823  (13\%)  & 1132  (4\%)   & 7081  (25\%)   & 7283  (26\%)   & 9057  (32\%)     \\
W4     & 1825  (6\%)   & 3991  (14\%)  & 1109  (4\%)   & 6519  (23\%)   & 7112  (25\%)   & 9310  (33\%)     \\
\hline  
epoch 1&  1983 (9\%) & 3668 (19\%)&  1091 (6\%)   &   5086 (26\%) &  4785 (25\%) & 4579 (24\%)    \\
epoch 2&  1705 (4\%)  & 4146 (11\%)&  1150 (3\%)  &   8514 (23\%)&  9610  (26\%)& 13788 (37\%)   \\   
\hline           
total &  3688 (6\%)& 7814    (14\%)&  2241   (4\%) & 13600   (24\%) &  14395   (26\%)& 18367  (33\%)    \\
%\hline  
%epoch 2a&  789  (8\%) &  975 (10\%)&       (\%)   &   2044 (21\%) &  2583 (26\%) & 3573 (36\%)    \\
%epoch 2b&  2864 (10\%)  & 3183 (11\%)&       (\%)  &   6442 (22\%)&  7001  (24\%)& 10178(35\%)   \\   
\end{tabular}
\end{table*}
% flag percentage

\begin{figure}
\resizebox{\hsize}{!}{\includegraphics[clip=true]{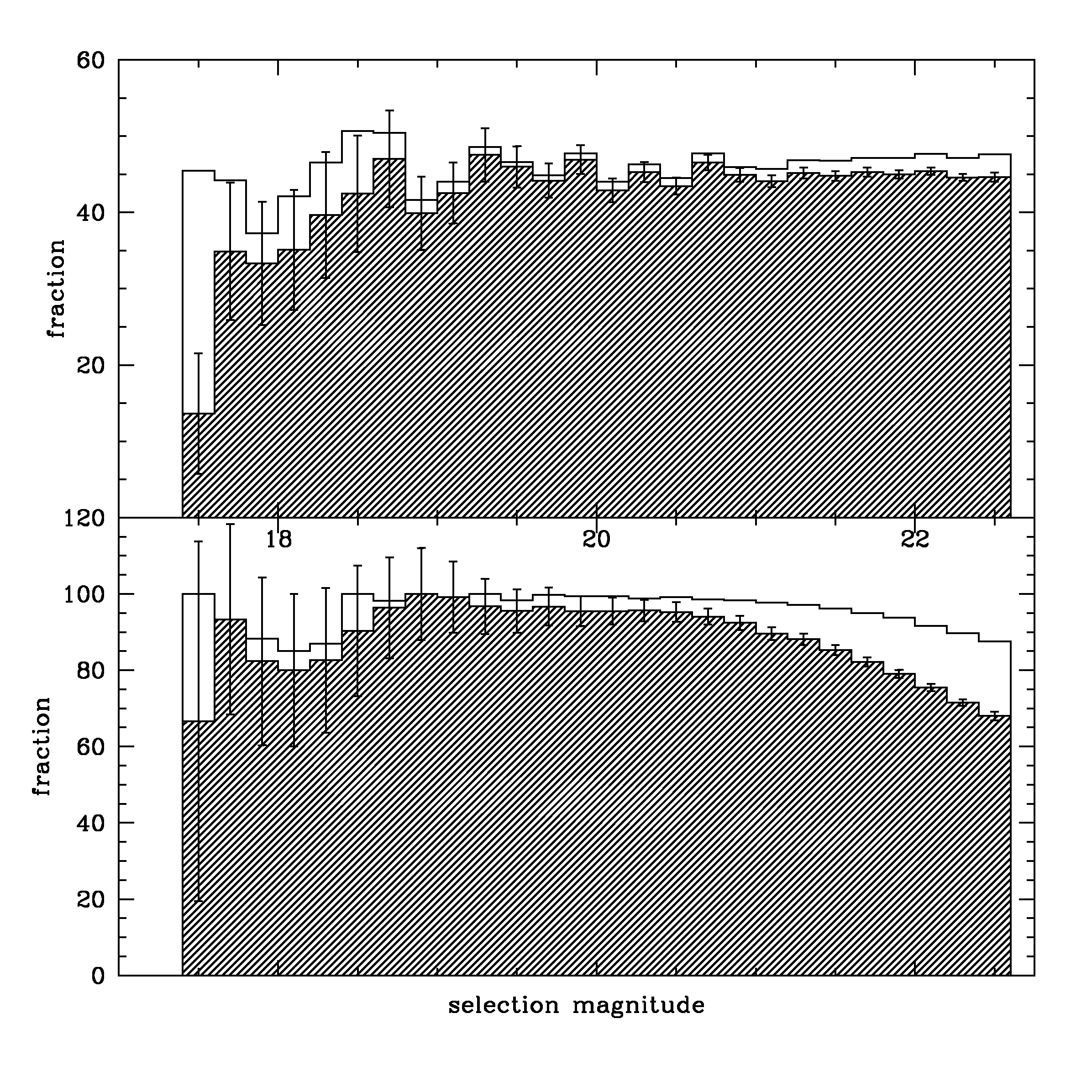} }
\caption{
Top: observed target sampling rate (empty histogram), and detection rate (shaded histogram) for the VIPERS survey targets. Bottom:
redshift measurement success rate (empty histogram), and reliable
measurement success rate (shaded histogram) targets. Error bars are
indicated for shaded histogram only for clarity.
}
\label{successRate}
\end{figure}

Figure \ref{successRate} shows the overall sampling of the VIPERS survey as a function of magnitude. 
The observed target sampling rate (top panel, empty histogram) is remarkably stable,
demonstrating that our target selection criterion is not flux dependent. Fainter than $i_{\rm AB}$ the detection rate 
(top panel, shaded histogram) is constant around 45\%. The slight decrease we observe at brighter magnitudes
is probably due to the short slits affecting the extraction of the brighter objects. We emphasise
that such a decrease remains within the uncertaintiesm but for the very first magnitude bin and we are currently
revising the spectra extraction procedure to ameliorate this effect.    
Conversely, there is a clear (albeit small) trend for redshift measurement success rate to diminish with increasing 
object flux(bottom panel, empty histogram): in the last half-magnitude bin the fraction of objects for which a redshift 
measurement is available drops from 93\%
to 88\%. This drop becomes even more dramatic when only reliable measurements are considered (bottom panel, shaded histogram). 
In section  \ref{weightsAndMask} we will illustrate how 
it is possible to account for these effects statistically when using the VIPERS spectroscopic sample.
  
% measurement vs. mag

\subsection{Redshift distribution}
% z distribution:ALL SECURE/UNSECURE, GALAXIES SECURE/UNSECURE, AGN SECURE/UNSECURE
% macro zdist3.mac
\begin{figure}
\centering
\includegraphics[clip=true]{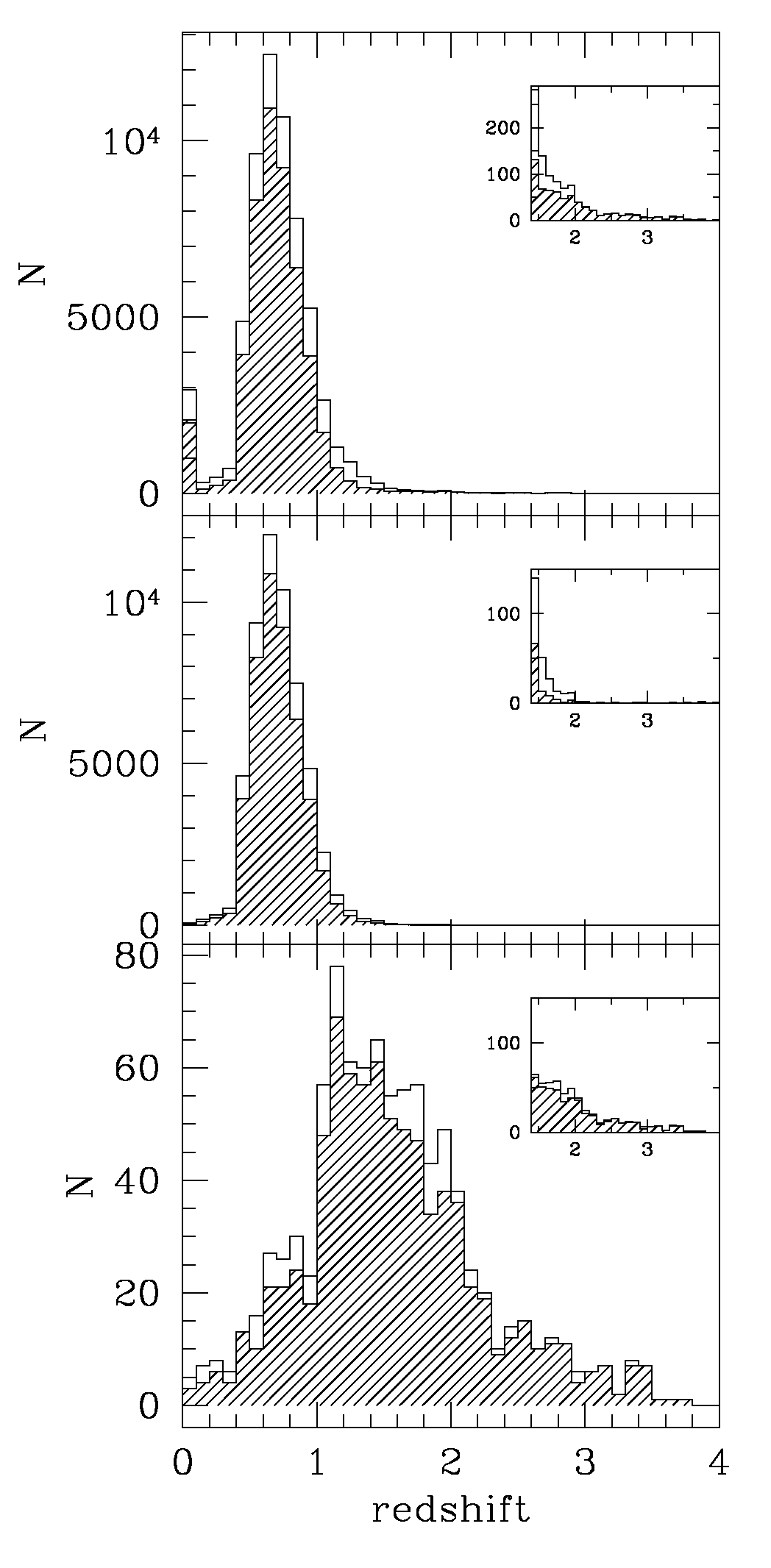} 
\caption{Redshift distribution for the VIPERS survey Top: all objects, middle: only galaxies; bottom: only AGN. 
The smaller insets zoom
on the high redshift tail. The histograms include secondary objects. Empty histograms are for all measured objects, shaded
histogram for reliably measured redshifts.}
\label{zdist}
\end{figure}
%
%data success rate table
% secure include 9.x flags
% to be revised after AGN checks
\begin{table}
\caption{VIPERS measured redshift for stars galaxies and AGN}
\label{summaryTable}
\centering
\begin{tabular}{c c c c c c c c}
\hline\hline
Field   & galaxies &reliable          & stars & AGN & reliable      \\
        &          &       galaxies &       &     &   AGN \\
\hline                          
W1      &   27336  & 23662          &  598  &  442 & 380  \\
W4      &   26010  & 22196          &  1648 &  383 & 338 \\
second objects & 581 & 371          &  202  &  4   & 3   \\
total   &   53927  & 46229          & 2448  & 829 & 721    \\
\end{tabular}
\end{table}
Table \ref{summaryTable} shows the number of target and serendipitous stars, galaxies and AGN, considering all measurements or
only reliable (flags 9.x,2.x,3.x,4.x) redshifts.
The stellar contamination is impressively low ($\sim~4\%$), demonstrating that the adopted criterion for excluding stars from the sample works well. 
The higher number of stars in the W4 area reflects the lower galactic latitude of this field with respect
to W1 ($-44$ degrees and $-57$ degrees respectively). 
Figure \ref{zdist} shows the redshift distribution for the whole VIPERS sample (top panel), the galaxy sample (middle panel)
and the AGN sample (bottom panel). 
The observed redshift distribution shows that our selection criterion works
well in excluding low redshift galaxies. As expected, the highest
redshift tail is made up of AGN, while only a handful of galaxies 
have a redshift above 1.2.
\\
\section{Survey selection function: masks and weights}
\label{weightsAndMask}
The survey selection function quantifies the actual probability
that a galaxy with given properties at a given position on the sky is
actually observed by VIPERS.  This can be a binary probability, as implied
by the global survey geometry and instrumental footprint, or a more
complex function of position on the sky (as e.g. introduced by a varying
sampling rate due to the limited number of slits that can be
accommodated in each quadrant).  In more detail, it can include a
continuous function that depends in general on
the galaxy flux, colour, spectral properties and redshift.  The
accurate knowledge of the actual selection function resulting from all
these effects allows us to define a set of weights to be applied to each galaxy.
These are used to
renormalize the observed density and two-point statistics to those one
would ideally obtain from a statistically complete sample. 
In this section we shall describe in some detail how the different contributions
have been estimated and encapsulated into a set of masks and weights
describing the VIPERS selection function, which are provided to the general
user together with the data release. These same masks and
weights have been used, for example, in the 
estimate of the VIPERS mass and luminosity functions
\citep{vipers_mf, vipers_cm} and in the clustering studies of
\citet{vipers_clus} and \citet{vipers_clus_lum}, where further details
can be found.
\\
\subsection{Photometric and Spectroscopic Masks}
\label{masks}
First, we consider selection effects arising from the CFHTLS parent
catalogue.  The photometric survey is not free from defects and there
exist a small number of gaps due to failed observations and artefacts
associated with bright stars that affect the VIPERS area.  These
regions are treated as holes in the survey and are described by a
photometric mask.

The area surveyed by the spectroscopic observations 
may be described by the geometry of the
VIMOS footprint and the lay-out of the observed pointings.  Although the
instrument geometry is fixed from night to night, particular observing
conditions can modify the effective area.  For instance, in certain
observing configurations, a VIMOS quadrant may be partially obscured
by the telescope guider arm.  We have reconstructed the precise geometry of
each quadrant and recorded it as a polygon.  
The combination of such polygons defines a second binary mask, which allows
a precise reconstruction of the overall survey footprint on the sky.
The construction of the photometric
and survey masks is described in detail in \citet{vipers_main}.

The application of the two masks to the whole target catalogue (see Section 
\ref{design}) defines what in the following we call the parent sample.

\subsection{Target Sampling Rate}
\label{TSR}
Only $\simeq 45$\% of the available targets in the parent sample may
be assigned a slit and observed according to the slit-assignment
strategy.  The galaxy Target Sampling Rate (TSR)
is defined as the fraction of candidate galaxies (according to our selection criterion, see Section~\ref{design})  in the parent sample
for which a spectrum has been acquired. Note that this is slightly different
from the target detection rate shown in Figure \ref{successRate} and Table
\ref{detectionTable}, where also AGN candidates were considered.
The TSR is both a function of angular position on the sky and apparent flux. 

As for all multi-object spectrographs, the TSR varies with location on
the sky, depending on the fluctuations in the surface density of objects. 
In particular, the single pass strategy of VIPERS generates a 
{\it proximity bias}, in which galaxy close pairs are disfavoured.
This is particularly relevant in two-point
statistical studies, 
e.g. for two-point correlation functions, for which a specific
correction based on the angular clustering of the measured and parent samples
has to be applied \citep{vipers_clus}. The TSR as defined here (indicated as TSR(Q)
in the following) maps the quadrant to quadrant variations in 
the target sampling.

Given the way we have built the spectroscopic masks (see section~\ref{design}), the TSR is expected to be independent of the target
apparent selection magnitude. Still, a very slight magnitude dependence exists at very bright magnitudes
(see Figure \ref{successRate}, top
panel).
This can be due either to the higher
angular clustering of the brightest objects, or to the short slits 
which may
disfavour the very largest objects (see Section \ref{observations}), or to a
combination of both effects.
The TSR used in published papers (and distributed as part of this release) 
takes this effect into account: see \citet{vipers_mf, vipers_cm} for details.

\subsection{Spectroscopic Success Rate}
\label{SSR}
The Spectroscopic Success Rate (SSR) is defined as  the fraction of
measured candidate galaxy targets with a reliable (flag 2,3,4,9) galaxy redshift measurement over all the detected 
galaxy targets. 
As for the TSR, this is slightly different from the reliable measurements success rate
shown in Figure \ref{successRate} and Table \ref{detectionTable}, where we
have included also AGN candidates.
The SSR is
clearly sensitive to both the observing conditions and the apparent magnitude.
A second order dependence on spectral type and redshift is usually present, 
as, at a given magnitude, it can be easier 
to measure redshifts for strong emission line galaxies, 
if the emission lines fall within the observed wavelength range.  

In the clustering analysis of \citet{vipers_clus}, the angular dependence of the
SSR (SSR(Q)) was estimated as a function of VIMOS quadrant as 
the ratio between the number of reliable
redshift (flag $2.\ast\le z_{\rm flag}\le
9.\ast$) and the total number of measured targets.
The average SSR(Q)
value is typically $>80$\%, falling to $50$\% for quadrants
observed in particularly poor conditions.

Note that, in principle, one can expect angular variations of the TSR and SSR 
on scales smaller than those of 
a single quadrant, for example due to the proximity bias discussed in the previous section or to an imperfect centring of the objects in the slit due to 
optical distortions. 
These effects cannot be viewed purely as a
position-dependent probability of obtaining a redshift. This means that
for some specific analyses, a more sophisticated treatment of these
effects may be required, as discussed in \citet{vipers_clus}.
The various information we provide within PDR-1 should in general allow
such higher level of refinement, if required.

In the case of statistical analyses which are not sensitive to
  the angular position of objects, it is possible to disentangle the dependence
of SSR on apparent magnitude and redshift using the full VIPERS sample and use only the SSR(mag,$z$) 
completeness weights \citep{vipers_mf, vipers_cm}. 
The SSR(mag,$z$) has been estimated by computing 
the  ratio of the number of successful galaxy
redshift to the total number of detected galaxy targets per magnitude and redshift bins. However, it is important to note
  that the SSR(mag,$z$) corresponds to the SSR averaged over all
  quadrants and cannot be used simultaneously with the SSR(Q)
  previously defined, which instead corresponds to the averaged SSR
  over the magnitude and redshift distributions in a given quadrant.
 
\subsection{Colour Sampling Rate}
\label{CSR}
The Colour Sampling Rate (CSR) defines the completeness of
   galaxies detected in spectroscopy with respect to a purely $i'<22.5$
  magnitude-limited sample. It accounts for the missed galaxies at the
  boundaries of the colour-based redshift selection criterion.
Such
incompleteness affects only galaxies with redshift close to the
nominal lower limit of $z\simeq 0.5$.
The adopted colour criteria were in fact calibrated using the 
purely flux limited data of the
VVDS survey \citep{vipers_main}. These calibration tests show that in the
$(r-i)$ vs $(u-g)$ plane, the distance of a galaxy 
from the adopted threshold is a monotonic function of redshift.
As such, the CSR is expected to be close to a step function 
changing from 0 to 1 around the nominal redshift threshold of $z=0.5$.

\begin{figure}
\resizebox{\hsize}{!}{\includegraphics[clip=true]{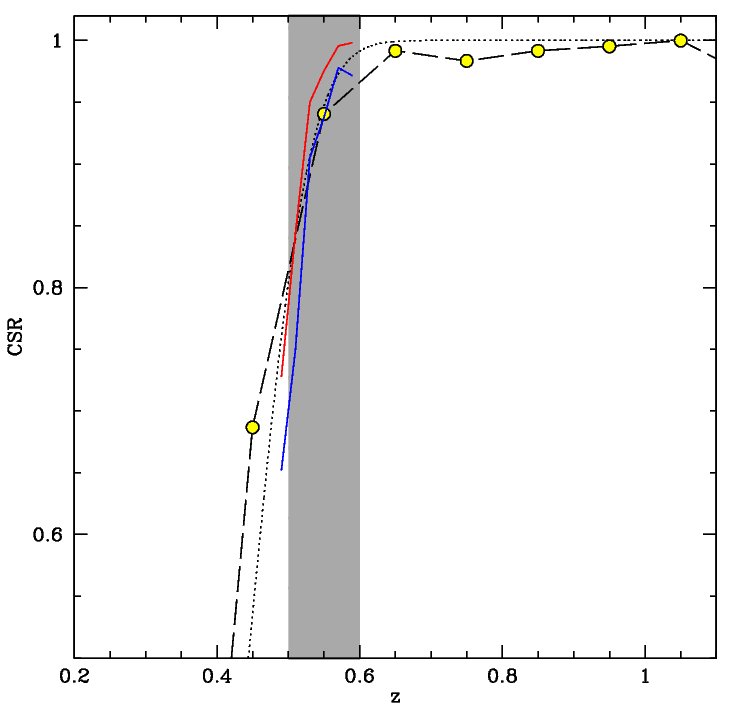}}
\caption{Colour Sampling Rate (CSR) for the VIPERS spectroscopic
  sample. Dashed line and dots: the CSR as computed from the VVDS sample; 
dotted line: the functional fit; 
 solid lines: CSR computed from the photometric sample for blue and red galaxies.
The shaded area underlines the redshift range where the CSR has to be applied.}
\label{CSRFig}
\end{figure}

To compute the CSR, we have used the VVDS Deep and
Wide datasets, which not only are purely flux limited surveys but also  
share the same CFHTLS photometry of VIPERS. 
%{\bf In the computation, we have focused on the selection boundaries (between 0.5 and 1.2 in redshift), as 
%trying to correct for the selection function beyond the limits of the selection function itself would
%be prone to too many uncertainties.}
The result is shown in Figure ~\ref{CSRFig}.  
The measured CSR points are well described by 
\begin{equation}
\label{CSReq}
{\rm  CSR}(z)=0.5-0.5\,{\rm erf}[b (z_{\rm t}-z)],
\end{equation}
where ${\rm erf}$ is the
error function, $b = 10.8$ and $z_{\rm t}=0.44$.  
Compared to the VVDS data points, this fitting
function only slightly overestimates the CSR (and therefore underestimates
the weight) for galaxies at redshift $z=[0.55,0.6]$. 
Comparison of the CSR obtained in this way to that 
obtained using photometric redshifts
(therefore using a much larger number of galaxies), 
shows a very good agreement, as
visible in Figure ~\ref{CSRFig}.
%from  micol's mail,24 sept. at 17:52 
We also checked whether passive
and active galaxies show a different CSR, by splitting the sample
on the basis of either the $NUVrK$
diagram or the $U-B$ bimodality \citep{vipers_cm}. As no significant difference has
been found, we concluded that the only relevant dependence of the CSR is the
one on redshift.

The final
weight to be assigned to a given galaxy will be the product of the
three weights mentioned above, i.e. 
$$w={\rm TSR}^{-1} \cdot {\rm
  SSR}^{-1} \cdot {\rm CSR}^{-1} = w_{\rm TSR}\cdot w_{\rm SSR}\cdot w_{\rm CSR}$$
In the case of angular position-dependent
  statistics, such as the two-point correlation function, the
  primary weights to be used are the TSR(Q) and SSR(Q) and the other
  dependences and possible corrections have to be studied in more
  details on a case-by-case basis (see \citet{vipers_clus}).
\\
In Section~\ref{database} we explain how to get access to the
survey masks and weights.
\\
\section{VIPERS spectra}
\begin{figure*}
\resizebox{\hsize}{!}{\includegraphics[clip=true]{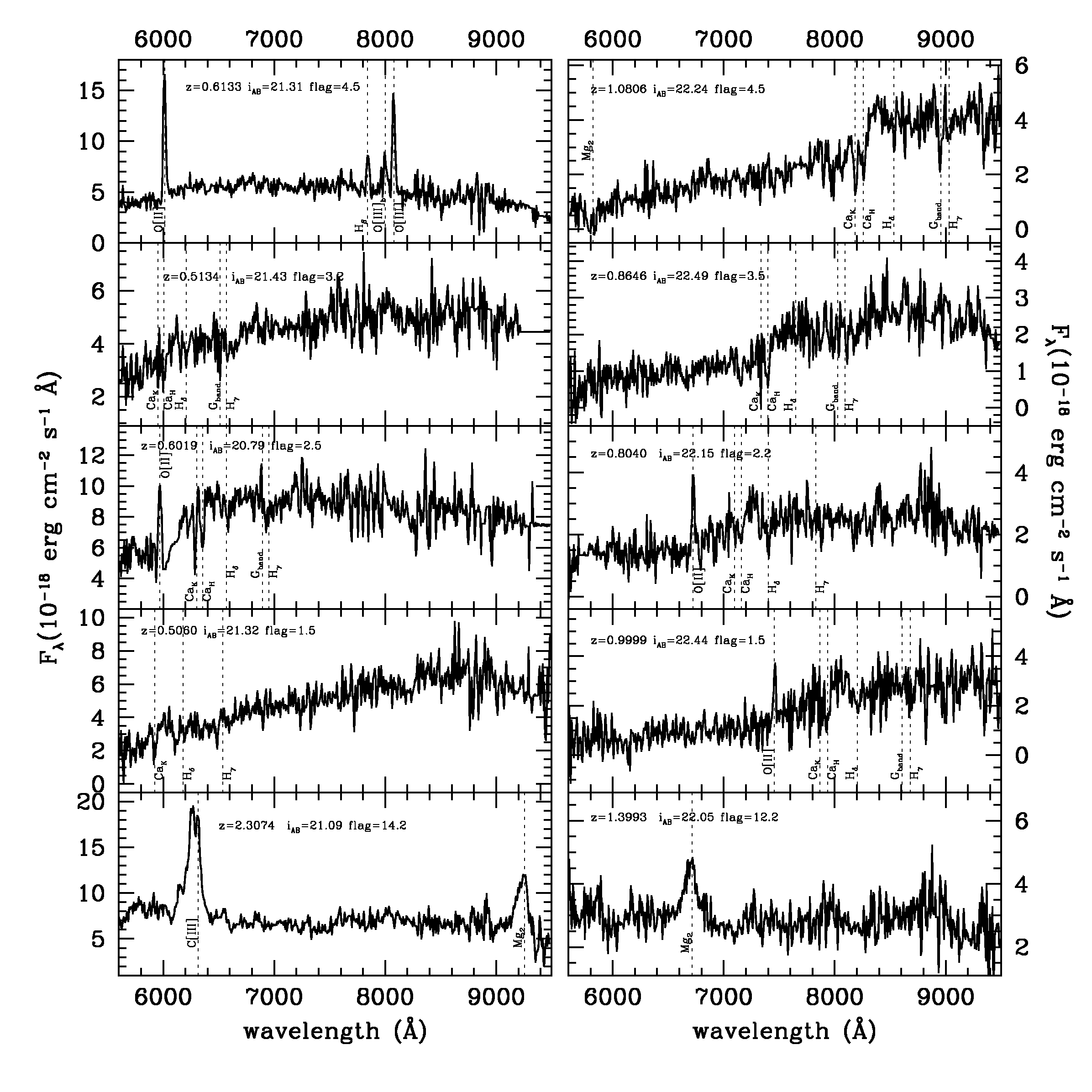}}
\caption{Some representative VIPERS spectra: first column for mid-brightness objects, right column for 
objects close to the survey magnitude limit ($i_{\rm AB}=22.5$). Redshift reliability flags range from highest reliable (flag 4.5, first row) to
least reliable (flag 1.5, 4$^{\rm th}$ row). In the last row we show two AGN spectra. }
\label{spectra}
\end{figure*}

Figure \ref{spectra} shows a few examples of VIPERS spectra, for
galaxies with different redshift and reliability flag. 
All spectra have been normalized to the object $i_{\rm AB}$ magnitude (see Section \ref{reduction}). 
\subsection{Spectral feature measurements}
Given the redshift distribution and the wavelength coverage of the VIPERS survey, the strongest 
spectral features which can be measured are the Balmer break and the most
prominent emission lines, namely [OII], [OIII] doublet, [H$\beta$] and [H$\alpha$].
The Balmer break (D4000$_n$) has been computed as  the ratio between the mean flux measured over the 
4000-4100\AA~ range and the 
3850-3950\AA~range \citep{D4000}. The
error on the D4000$_n$ break is computed by propagating the error on the two means. 
Measurement of line fluxes can be performed in various ways: 
from the simplest  pure flux integration below the line, to the most sophisticated approaches,
which attempt to take into account the absorption component of the Balmer lines.
The latter approach is particularly indicated when the purpose is to
derive a precise estimate of the gas component, for example
in all studies involving metallicity measurements. Our first aim is to separate emission line galaxies from
non emission line galaxies, and for this reason we have chosen the simplest approach of pure 
flux integration below the line.
All computations are done in counts and converted into fluxes using the counts-to-flux conversion factor 
derived from the instrument sensitivity function. For each line, a local continuum per pixel unit is computed as the mean of the counts in 
two regions redwards and bluewards of the line :
\begin{equation}
{\rm C}_{pix}=\frac{\sum_{i \in {\rm NcontPix}}{(C_i)}}{{\rm NcontPix}}.
\end{equation}
where $C_i$ are the counts in pixel i and  ${\rm NcontPix}$ are the pixels 
within the blue and red wavelength ranges used for the continuum
computation.
The error on the continuum level  is computed using the noise spectrum  associated to each source
(see equation \ref{noiseSpectrum}):
\beq
{\sigma_{Cpix}}=\frac{\sqrt{s^2 +{\rm <skyResidual>}^2}}{\sqrt{\rm NcontPix}},
\eeq
where $s^2$ is the continuum variance.
The line counts are obtained integrating the continuum-subtracted counts in the defined line region
  \beq{\rm LineCounts}=\sum_{i \in {\rm lineRegion}}{(C_i-{\rm C_{pix}})} \eeq
and their error 
  \beq\sigma_{LC}=\sqrt{\sum_{i \in {\rm lineRegion}}{C_i}+NL^2*\sigma_{Cpix}^2} \eeq
where $NL$ are the pixels within 
the line region.
Table \ref{lines} gives the line and continuum boundaries we have used for the different features we have measured.
% line measure
\begin{table}
\caption{Spectral feature parameters.}
\label{lines}
\centering
\begin{tabular}{c c c c }
\hline\hline
Spectral  & Line    & Continuum   &  Continuum  \\
feature   & Range  &  Range (blue)   &  Range (red)  \\
\hline
$[{\rm OII}]$   &  3710-3745    &3600-3700&	3755-4000\\
$[{\rm OIII}]_a$ & 4990-5025	&4700-4825&	5035-5150\\
$[{\rm OIII}]_b$ & 4900-4990	&4700-4825&	5035-5150\\
${\rm H}_\beta$  & 4845-4880	&4700-4825&	5035-5150\\
${\rm H}_\alpha$ & 6535-6600	&6350-6500&	6610-6680\\
\end{tabular}
\end{table}

\subsection{Spectroscopic properties}
% D4000 distribution: macro D4000.mac in dir. stacked
\begin{figure}
\resizebox{\hsize}{!}{\includegraphics[clip=true]{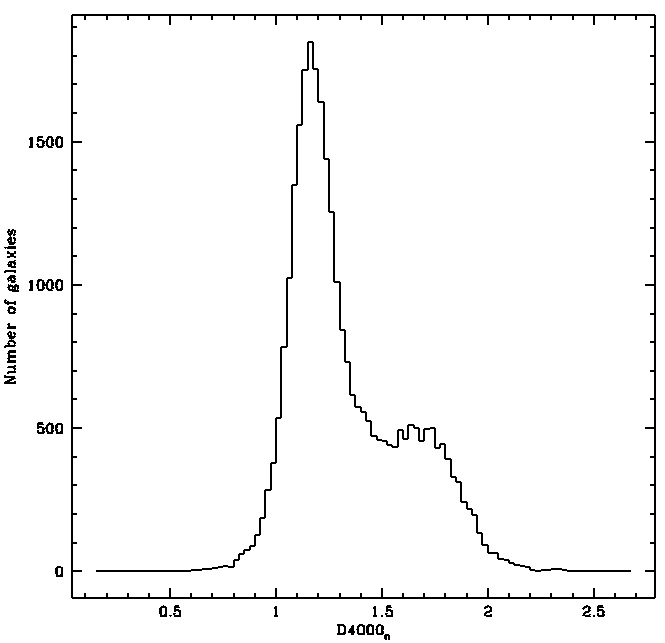} }
\caption{Balmer break distribution for the subsample of flag 3 and 4 VIPERS galaxies.}
\label{d4000Distribution}
\end{figure}
%
% D4000 O2: topcat font 19 bold
\begin{figure}
\resizebox{\hsize}{!}{\includegraphics[clip=true]{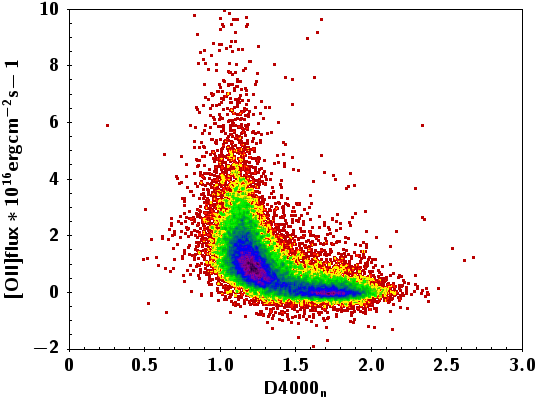} }
\caption{[OII] line flux vs. Balmer break amplitude density plot for the subsample of flag 3 and 4 VIPERS galaxies.
Colour coding ranges from black  to red with decreasing density.}
\label{O2D4}
\end{figure}
Spectroscopic information can be used to subdivide
galaxies into different classes.
In Figure \ref{d4000Distribution} we show the distribution of the D4000$_n$ break amplitude
for the subsample of
flag 3 and 4 galaxies: its bimodality is evident.
This figure is qualitatively identical to the bimodality plots presented in \citet{vipers_cm}, 
which are based on the galaxy rest-frame $U-V$ colour.
In Figure \ref{O2D4} we plot the [OII] flux as a function 
of the D4000$_n$ break value for the same galaxy subset. 
In this Figure, the division into two large groups of galaxies is even more evident: most of the galaxies
with a D4000$_n$ break larger than 1.5 have no or little sign of emission lines (or on-going star formation), while
galaxies with strong on-going star formation are younger (i.e. have a lower D4000$_n$ break).
A deeper classification of galaxies on the basis of spectral properties will
be the subject of future work; here
we restrict ourselves to investigating how the spectral features listed in Table \ref{lines} relate to
other classification schemes used within the VIPERS project.

% D4000 O2 SED: topcat
\begin{figure}
\resizebox{\hsize}{!}{\includegraphics[clip=true]{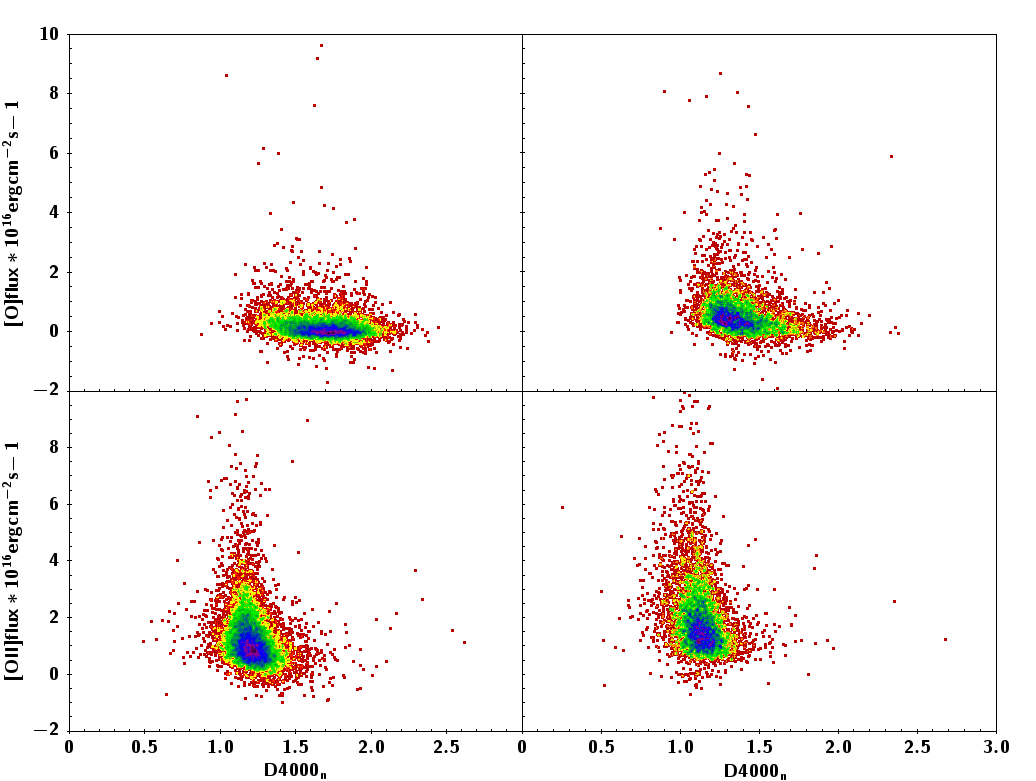} }
%{\includegraphics[width=0.5\columnwidth,keepaspectratio,clip=true]{D4O2SED1.png}}{\includegraphics[width=0.5\columnwidth,keepaspectratio,clip=true]{D4O2SED2.png}}\\
%{\includegraphics[width=0.5\columnwidth,keepaspectratio,clip=true]{D4O2SED3.png}}{\includegraphics[width=0.5\columnwidth,keepaspectratio,clip=true]{D4O2SED4.png}}
\caption{[OII] line flux vs. Balmer break amplitude for the subsample of flag 3 and 4 VIPERS galaxies by SED type: 
type 1,top left; type 2, top right, type 3 bottom left, type 4 bottom right. Colour coding as in figure \ref{O2D4}}
\label{O2D4SED}
\end{figure}

In \citet{vipers_mf} an   
SED-based classification was proposed: according to the best fitting template, galaxies were divided into 
four SED types, where type 1
are red, old and supposedly quiescent galaxies, while type 4 are the bluest and most active
galaxies. In \citet{vipers_cm} it is shown how, using this classification, type 1 galaxies
define fairly well the red sequence in the colour magnitude diagram. 
We would expect type 1 galaxies to show no (or very little) sign of star formation activity. 
Figure \ref{O2D4SED} shows how the different SED type galaxies cluster in the $D4000_{n}-$[OII] flux plane.
Although on average type 1 galaxies show no sign of star formation, for 
30\% of them we detect significant (at least $3\sigma$) [OII] emission. Conversely, type 4 galaxies totally 
dominate the high [OII] flux tail, and never show a prominent Balmer break.
In other words, the SED based classification is rather efficient in isolating
highly star forming galaxies (type 4), but the earlier type galaxies thus selected are not necessarily quiescent.

% D4000 O2 PCA:
\begin{figure}
\resizebox{\hsize}{!}{\includegraphics[clip=true]{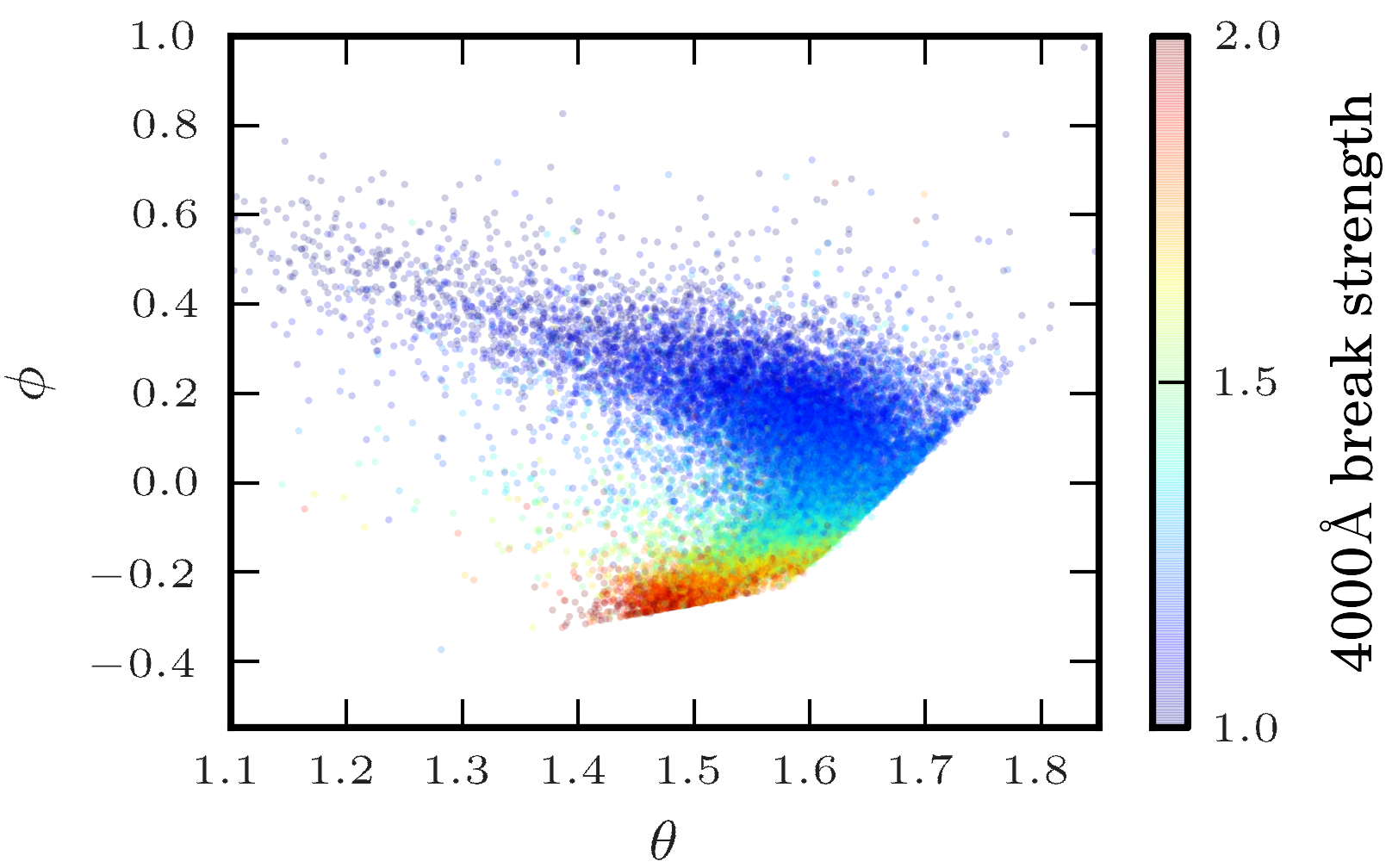} }
\caption{The PCA galaxy classification plane, with Balmer break amplitude shown in colours.}
\label{D4PCAGroup}
\end{figure}
% D4000 O2 PCA:
\begin{figure}
\resizebox{\hsize}{!}{\includegraphics[clip=true]{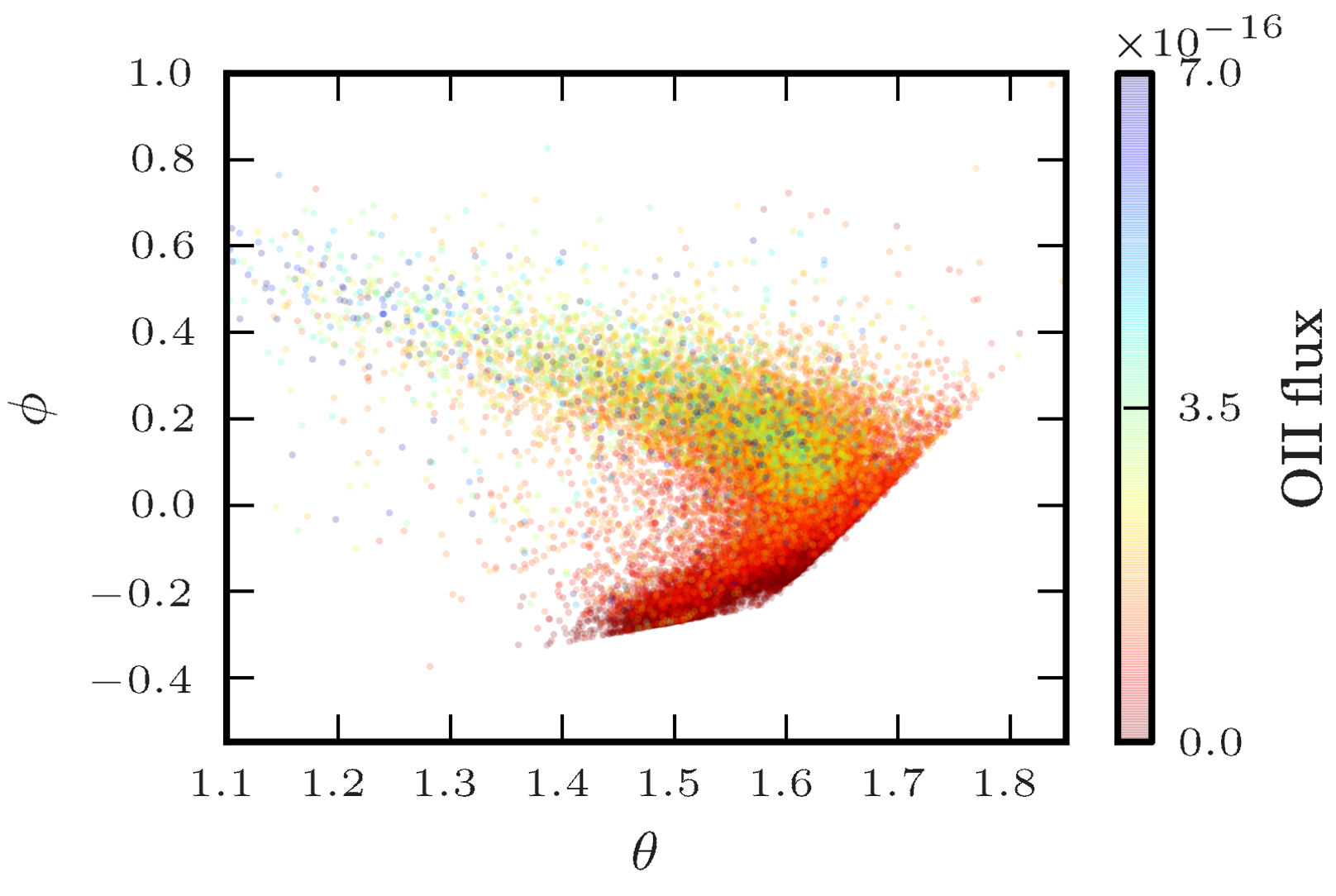} }
\caption{The PCA galaxy classification plane, with [OII] line flux shown in colours.}
\label{O2PCAGroup}
\end{figure}
A different approach was taken in \citet{vipers_pca}, where a Principal 
Component Analysis has been used to derive the first three eigen-coefficients, and
a group-finding algorithm has been  applied on the
resulting $\theta - \phi$ diagram, dividing galaxies into 15 different groups: groups from 1 to 8
comprise the classical types from E to Sc, while group from 9 to 15 comprise the more active starburst galaxies.
In Figure \ref{D4PCAGroup} we show the PCA $\theta - \phi$ diagram with galaxies 
in different colours according to their value of Balmer 
break. We can see that for galaxies showing $D4000_{n}$ higher than 1.5, the break
amplitude is well correlated with the $\phi$ parameter.
This is consistent to
the classification proposed in \citet{vipers_pca}, who indicate this part of the diagram 
as beeing populated by the earlier galaxy types. 
Figure \ref{O2PCAGroup} relates the PCA $\theta - \phi$ parameters with the flux of the [OII] emission line.
The strongest emission line galaxies populate the locus of the starburst galaxies,
as identified by the PCA classification. Galaxies in the locus of mid and late spirals ($\theta>1.6$ and $\phi >0 $)
do not seem to be strong [OII] emitters. This could partly be due to 
the limit on [OII] flux detection, and a more accurate analysis taking into account
line detection  upper limits will be done.
Overall, there is a good agreement between the PCA based classification and the 
measured spectral features in identifying the extreme galaxies (early types or Starbusts). 
This is not surprising, as PCA itself is based on spectral decomposition, 
and confirms the validity of the PCA approach. 
However we note that a one to one correspondance between
PCA parameters and spectral features would not be appropriate, 
as the PCA takes into account the full spectral shape. Neither line flux nor Balmer break alone can
predict the overall SED accuracy as PCA can do
\\
It is beyond the scope of this paper to attempt a galaxy classification 
scheme based solely on spectral features.
In a forthcoming paper we will further
develop the subject of galaxy classification, quantitativey comparing the
different possible schemes.

\subsection{Global spectral properties}
\label{stacked}
\begin{figure*}
\resizebox{\hsize}{!}{\includegraphics[clip=true]{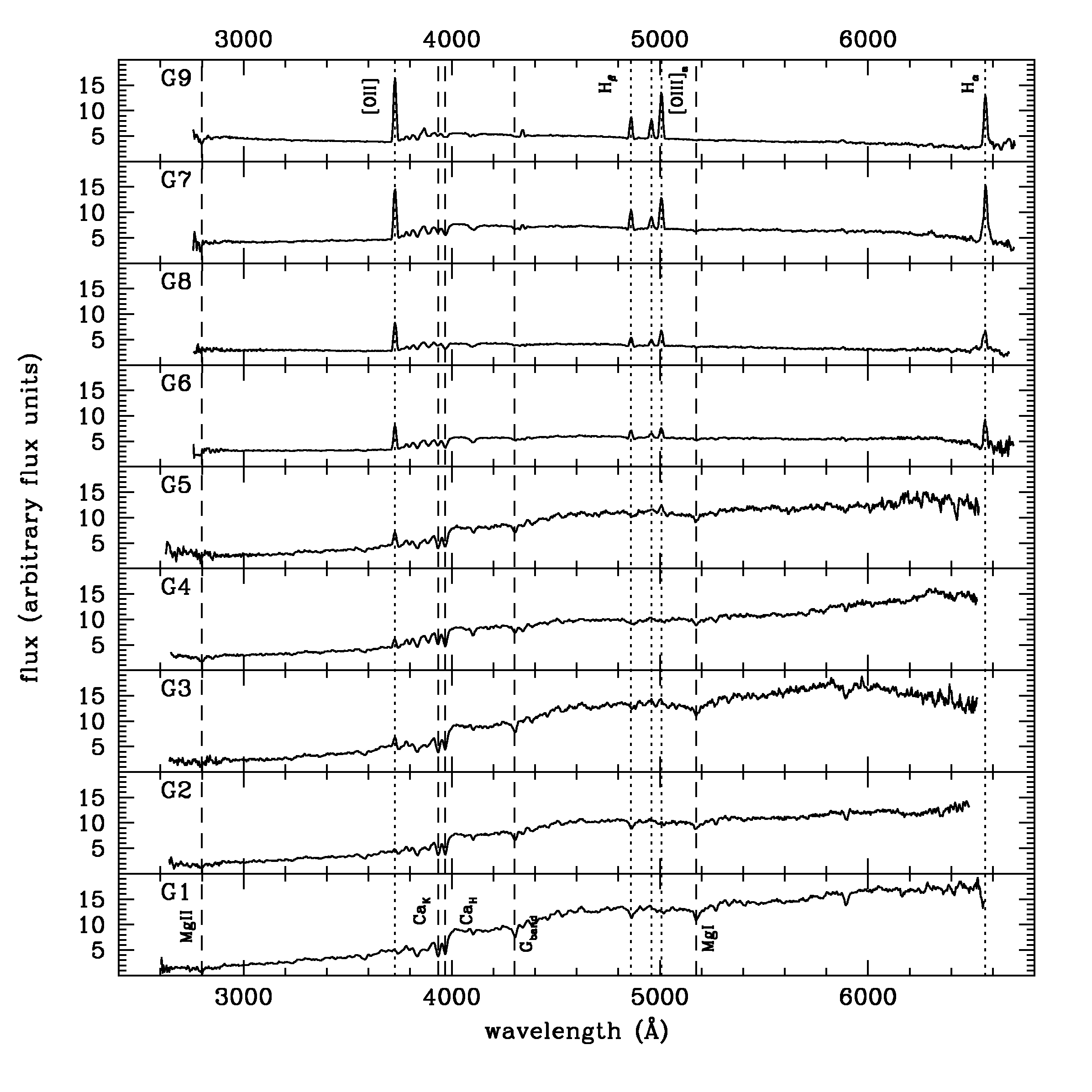}}
\caption{VIPERS rest-frame stacked spectra for different galaxy groups according to spectral feature strength. Vertical 
lines indicate location of the strongest features, dotted for emission and dashed for absorption lines.}
\label{stackedFig}
\end{figure*}

\begin{table}
\caption{Spectral feature parameters }
\label{stackedTab}
\centering
\begin{tabular}{c c c c }
\hline\hline
Group  & $D4000_{n}$  & Line Flux$^1$ & N. of  \\
name   & Range        &  Range       & galaxies\\
\hline
G1     & >1.7         & undetected  &   3457      \\
G2     & 1.55-1.7     & undetected  &   1882      \\
G3     & >1.7         & $>3.0\times10^{-17}$ &    1135     \\
G4     & 1.15-1.55    &  undetected &    2539     \\
G5     & 1.55-1.7     & $>3.0\times10^{-17}$ &   963       \\
G6     & 1.15-1.55    & $<1.0\times10^{-16}$ &  6198        \\
G7     & <1.15        & $<1.0\times10^{-16}$ &  1123        \\
G8     & 1.15-1.55    & $>1.0\times10^{-16}$ &  6681        \\
G9     & <1.15        & $>1.0\times10^{-16}$ &  4817       \\
\hline
\end{tabular}
\begin{tabular}{l c}
$^1 \rm erg\,cm^{-2} s^{-1}$ 
&~~~~~~~~~~~~~~~~~~~~~~~~~~~~~~~~~~~~~~~~~~~~~~~~~~~~~~~~~~~~~~~~~~~~~~~~~~~~~~~~~~~~~~~~~~~~~~~~~~~~~~\\
\end{tabular}
\end{table}
The VIPERS spectral resolution $R\sim250$  allows us to
study individual spectroscopic properties only for galaxies with the highest signal to noise ratio. On the other hand,
the high statistics provided by the VIPERS galaxy sample allows 
to define different galaxy groups (according to the scientific problem one may want to address),
make high quality stacked spectra for each group and perform spectral measurement which would be
impossible on the single objects.
As an example,  we have divided VIPERS very reliable
galaxies (flag 3 and 4)  according
to the amplitude of the Balmer break and [OII] flux. Table \ref{stackedTab} 
shows the $D4000_{n}$ and [OII] flux limits we have used for each group, as well
as the number of spectra pertaining to that group.
We have stacked together the rest frame
spectra within each group, and the result is shown in Figure \ref{stackedFig}.
These kind of high quality observed spectra of galaxies at medium redshift 
can be useful in cross-correlation algorithms, in conjunctions with or alternative to
synthetic templates or the classical low redshift templates from \citet{kennicutt_atlas}, and
for this reason we include them in the distributed products.
More detailed studies of the spectroscopic properties of VIPERS galaxies making use of
stacked spectra are currently ongoing. 
\\
\section{Public Data Release and Database Access}
\label{database}
The data of the first VIPERS Public Data Release (PDR-1) are available at
{\tt http://vipers.inaf.it}. The release comprises:
\begin{itemize}
\item {\bf Spectroscopic catalogue}, described in Table~\ref{spectroCat}.
\item {\bf Parent photometric catalogue}, described in Table~\ref{photoCat}. 
It contains all sources in the CFHTLS catalogue
falling within the surveyed area and having  $i_{\rm AB} < 24.0$. 
\item {\bf Photometric masks}, as described in \citet{vipers_main}. 
The photometric mask marks regions of the VIPERS survey in which the
target selection may be affected by poor photometric quality in the 
parent CFHTLS catalogue.  These regions should be excluded from 
scientific analyses.  The fraction of the area lost is 2.4\% for 
the W1 field and 3.0\% for the W4 field.
The photometric mask is provided separately for the two VIPERS fields
W1 and W4.  The data files follow the DS9 Region format version 4.1
and use the polygon data structure.
\item {\bf Spectroscopic masks}, as described in \citet{vipers_main}. 
These masks, together with the photometric ones, should be used to determine the angular selection function of
the survey. The masks are provided as a set of DS9 Region files (version 4.1).
Each pointing in the survey consists of 4 quadrants  and
each quadrant is stored in a polygon data structure. Masks are provided 
for W1 and W4 separately, as well as for each single quadrant.
\item {\bf Quadrant dependent TSR and SSR} as described in 
Section~\ref{weightsAndMask}, provided in
the form of  two separate ASCII tables. In each table, the columns correspond to: pointing name, quadrant number, TSR (or SSR) value. The details on how these have been computed are given in \citet{vipers_clus}. The value of the TSR in each quadrant is the average
of the TSR over the apparent magnitude distribution within the 
quadrant.
\item {\bf Stacked spectra}, as shown in \ref{stacked}, in both ASCII and fits format.
\\
\end{itemize}
\begin{table*}
\caption{Spectral catalog contents }
\label{spectroCat}
\centering
\begin{tabularx}{\textwidth}{l X }
\hline\hline
Name  & Description   \\
\hline
id\_IAU     &   Object name, according to IAU standards: prefix VIPERS plus internal identification number       \\
num     & Internal id number (num) in the form  attxxxxxx where  a identifies the sky area (1 for W1 and 4 for W4), tt identifies the CFHTLS tile number where the object is located,  xxxxxx is the original CFHTLS ID within the tile      \\
alpha     & J2000 Righ Ascension in decimal degrees      \\
delta     & J2000 Declination in decimal degrees      \\
selmag     & $i_{AB}$ selection magnitude. The selection magnitude comes from CFHTLS T0005 catalogues   \\
errselmag     & Error on the selection magnitude       \\
zspec     & Spectroscopic redshift. A conventional zpsec value of 9.9999 is assigned in case a redshift could not be measured    \\
zflg     &  redshift confidence flag as described in Section~\ref{redshift}        \\
epoch     &  Observing epoch: epoch=1 objects have been observed before VIMOS refurbishing in summer 2010, epoch=2 objects have been observed after summer 2010      \\
photoMask   &  Flag indicating whether the object falls within the photometric mask. 1 if the object is inside the mask, 0 if it is outside         \\
tsr & Target Sampling Rate as described in Section~\ref{TSR}. TSR is = -1 for serendipitous targets (flags 2*.*), and observed objects not fulfilling the selection criterion for z>0.5 galaxies (i.e. AGN); TSR = -99 for objects outside the considered area and selection magnitude range (i.e. either i<17.5 or i>22.5)  \\
ssr & Spectroscopic Sampling Rate as described in Section~\ref{SSR}. SSR is = -1 for stars (z=0.0000), non measured objects (flag 0), low confidence redshift measurements (flags 1.*), spectroscopic BLAGN (flags 1*.*), serendipitous targets (flags=2*.*), and hight redshift galaxies (zspec>3); SSR = -99 for objects outside the considered area and selection magnitude range (i.e. either i<17.5 or i>22.5) \\
\hline
\end{tabularx}
\end{table*}

A detailed description of each quantity is provided in the distributed catalogues.
We note that with respect to the data used in previous VIPERS papers, in PDR-1 $\sim 0.5\%$ of the
redshifts have changed due to a deeper look into the more problematic spectra. As a consequence,
for this release we have re-computed also the TSR, SSR and CSR. However we underline that
these differences are too small to affect the scientific results contained in
previous papers.
\begin{table*}
\caption{Photometric catalog contents }
\label{photoCat}
\centering
\begin{tabularx}{\textwidth}{l X }
\hline\hline
Name  & Description   \\
id\_IAU     &   Object name, according to IAU standards: prefix VIPERS plus internal identification number.        \\
num     & Internal id number (num) in the form  attxxxxxx where  a identifies the sky area (1 for W1 and 4 for W4), tt identifies the CFHTLS tile number where the object is located,  xxxxxx is the original CFHTLS ID within the tile.       \\
alpha     & J2000 Righ Ascension in decimal degrees.     \\
delta     & J2000 Declination in decimal degrees.      \\
selmag     & $i_{AB}$ selection magnitude. The selection magnitude comes from CFHTLS T0005 catalogues.   \\
errselmag     & Error on the selection magnitude        \\
u,g,r,i,z & Magnitudes (AB system) from the CFHTLS T0005 catalogue, supplemented by T0006 catalogue in some specific cases (see \citet{vipers_main}, section 3 and appendix C, for details on the tile to tile color offests, as well as for T0005 and T0006 catalogue differences. All magnitudes are corrected for Galactic extinction.  When an object has not been observed in a given band, magnitude and error are set equal to -99. When a magnitude (and its error) could not be measured, these values are set to 99  \\
erru,errg,errr,erri,errz & Errors on CFHTLS05 magnitudes  \\
${\rm u_{T07},g_{T07},r_{T07},i_{T07},z_{T07}}$& Magnitudes (AB system) from the CFHTLS T0007 catalogue. All magnitudes are corrected for Galactic extinction  \\
${\rm erru_{T07},errg_{T07},errr_{T07},erri_{T07},errz_{T07}}$ & Errors on CFHTLS T0007 magnitudes.  \\
DeltaUG, DeltaGR, DeltaRI & Tile to tile color offsets used in the targets sample selection applied to the CFHTLS T0005 data (see \citet{vipers_main}, section 3.1)  \\
E$_{BV}$ & Extinction factor  E(B-V) derived from Schlegel's maps  \\
r2 & Radius enclosing half the object light as from CFHTLS T0005 catalogue,
in pixels  \\
r2$_{T07}$ & Radius enclosing half the object light as from CFHTLS T0007 catalogue,
measured in pixels  \\
classFlag & VIPERS selection flag based on the CFHTLS T0005 catalogue (see \citet{vipers_main}, section 4)  \\
agnFlag & A value equal to 1 is assigned to all AGN candidates, and equal to 0 otherwise (see Section~\ref{AGNselection}).  \\
photoMask   &  Flag indicating whether the object falls within the photometric mask: 1 if the object is inside the mask, 0 if it is outside    \\
spectroMask & Flag indicating whether the object falls within the spectroscopic mask: 1 if the object is inside the mask, 0 if it is outside (see Section ~\ref{masks}).\\
\hline
\end{tabularx}
\end{table*}

A full release of the one dimensional fully calibrated spectra, 
together with the spectroscopic features measurements, is foreseen in 2014.

Catalogues are distributed in three different forms: 
\begin{itemize}
\item A single tar file containing the full release
\item A web interface connected to  the VIPERS (see next section)
database (free registration required), which allows to perform SQL selections through an ergonomic, easy to use interface
\item At a later stage, data will also be queryable via the Virtual Observatory 
\end{itemize}

\subsection{VIPERS database web interface}
The WEB interface to the VIPERS data base allows one to perform queries 
 using a  confortable  User Interface. 
Once logged in, the user is presented with the list of tables to be queried. 
Queries can be applied to a single table, or to a combination of tables: 
if more than one table is selected, the query can be conducted 
either in parallel and independently on each of the tables, or
joining all tables, using the object ID as joining field.
Extensive information on the content of each table can be accessed by clicking on the table name, 
while generic help on the usage of the interface is available clicking on the HELP button on the left.
%\begin{figure*}
%\resizebox{\hsize}{!}{\includegraphics[clip=true]{TableSel.jpeg}
%}
%\caption{The VIPERS DataBase: Table selection form.}
%\label{TableSel}
%\end{figure*}

Queries can be of type  {\it Simple} or {\it Advanced}.
The {\it Simple} query  allows one to select all objects within a sky region (query {\it by Position}) or to 
retrieve objects within a user specified list (query {\it By Input List}),  or to
select objects satisfying simple selection conditions 
(query {\it By Parameters}).  
The {\it By Parameter} form allows one to set many selection conditions as ranges, 
equality or likeness, for example
%In Figure~\ref{query} we show an example 
%where we restrict the query to
extragalactic objects (zspec>0) with a secure redshift flag (zflg between 2 and 10) observed in epoch 2. 
Conditions can be imposed on any column of the selected tables and all the conditions are joined together
with an AND syntax.
The {\it Advanced} query panel page  has been thought to provide a more flexible tool for expert users. 
It allows to insert formulas in the selection statement, to define the selection condition in a customizable complex way, 
to sort the output using a specified data column order and it also allows one to limit the result output.

The query output page allows one to directly inspect the query results, to make simple simple plots for quick statistical inspection
of the results and to save results into ASCII files, VOTables or FITS binary tables.

% DB figures

%\begin{figure}
%\resizebox{\hsize}{!}{\includegraphics[clip=true]{query_small.png}
%}
%\caption{The VIPERS DataBase Interface: the {\it By Parameters} query form.}
%\label{query}
%\end{figure}

%
\section{Summary}
We have presented here the first first Public Data Release
(PDR-1) of the VIPERS survey, which includes 57\,204 spectroscopic measurements of
galaxies, AGN and stars.  Complementing the general description
given in \citet{vipers_main},  
we have discussed the details of the target selection, observations, data reduction and
redshift measurements, providing all relevant information for a proper use of the
data.  This includes the photometric and angular selection functions, 
as well as the weighting schemes to be adopted to correct for the survey incompleteness.

Both internal tests and comparison with external redshifts 
indicate a high reliability of the bulk of the VIPERS redshift data:
redshifts classified as Flags 3 and 4 (33\,102 objects)   %including secondary
in our internal grading system
show a confidence level of 99\% in their value, while that of the combined Flags 2
and 9 (49\,087 objects) %including secondary
is $>90\%$.   
Using more than 1500 objects
with repeated observations, we have estimated  
a redshift {\it rms} error of $0.00047(1+z)$ and shown that this value
does not vary significantly along the different Flag classes
considered as reliable (i.e. 2, 3, 4 and 9).  
The overall stellar contamination of the galaxy
target sample is found to be smaller than 3\%, confirming the goodness of our original star-galaxy
separation.

To provide hints on the global spectroscopic quality of the VIPERS
data,  we have presented first measurements of some basic spectral
features, as the Balmer break index and the strength of the
[OII]$~\lambda$3727 line.  We have also included first comparisons of these
spectroscopic markers of mean stellar age and star formation rate to galaxy
classification schemes used in other VIPERS papers. This clearly shows
the huge potential of this data set for statistical studies of the galaxy
population at an average redshift $z\sim 0.8$, an epoch where VIPERS
(already with this PDR-1), provides a significant leap in terms of
number of galaxies and volume. 

The full spectroscopic catalogues, together with the complementary
photometric information, survey masks and weights are publicly
available from  {\tt  http://vipers.inaf.it}. A full release of the
corresponding PDR-1 spectra is planned to take place in  2014. 

\begin{acknowledgements}
We acknowledge the crucial contribution of the ESO staff in the management of
service observations. In particular, we are deeply grateful to M. Hilker for his
constant help and support of this programme. We thanks an 
anonymous referee for the useful comments and suggestions.
Italian participation in VIPERS has
been funded by INAF through PRIN 2008 and 2010 programmes. LG and BRG acknowledge
support of the European Research Council through the Darklight ERC Advanced
Research Grant (\# 291521). OLF and LAMT acknowledge support of the European Research
Council through the EARLY ERC Advanced Research Grant (\# 268107). Polish
participants have been supported by the Polish National 
Science Centre (grants  N N203 51 29 38 and
2012/07/B/ST9/04425), the Polish-Swiss Astro Project (co-financed by a grant from
Switzerland, through the Swiss Contribution to the enlarged European Union), the
European Associated Laboratory Astrophysics Poland-France HECOLS and a Japan
Society for the Promotion of Science (JSPS) Postdoctoral Fellowship for Foreign
Researchers (P11802). GDL acknowledges financial support from the European
Research Council under the European Community's Seventh Framework Programme
(FP7/2007-2013)/ERC grant agreement n. 202781. WJP and RT acknowledge financial
support from the European Research Council under the European Community's
Seventh Framework Programme (FP7/2007-2013)/ERC grant agreement n. 202686. WJP
is also grateful for support from the UK Science and Technology Facilities
Council through the grant ST/I001204/1. EB, FM and LM acknowledge the support
from grants ASI-INAF I/023/12/0 and PRIN MIUR 2010-2011. YM acknowledges support
from CNRS/INSU (Institut National des Sciences de l’Univers) and the Programme
National Galaxies et Cosmologie (PNCG). CM is grateful for support from specific
project funding of the {\it Institut Universitaire de France} and the LABEX
OCEVU. The TOPCAT software \citep{topcat} has been widely use for this paper.
\end{acknowledgements}

\bibliographystyle{aa}
\bibliography{garilli_vipers_v2}
~\\
\end{document}